\documentclass[a4paper,10pt]{article}
\usepackage[a4paper, margin=1in]{geometry}  

\usepackage{amsmath}
\usepackage{microtype}
\usepackage{booktabs}
\usepackage{aas-macros}
\usepackage{float}
\usepackage{hyperref}
\usepackage[font=small,skip=0pt]{caption}
\usepackage{subcaption}
\usepackage[utf8]{inputenc}
\usepackage{graphicx} 
\usepackage{xcolor}
\usepackage{authblk}  
\usepackage[authoryear]{natbib}

\newcommand{\ShortName}{GalProTE} 

\title{\ShortName: Galactic Properties Mapping using Transformer Encoder}

\author[1]{Omar Anwar}
\author[1]{Brent Groves}
\author[1]{Luca Cortese}
\author[1]{Adam B. Watts}

\affil[1]{International Centre for Radio Astronomy Research (ICRAR), The University of Western Australia (UWA), 35 Stirling Hwy, Crawley WA 6009, Australia}
\affil[ ]{\texttt{omar.anwar@uwa.edu.au}}  

\date{March 2025}  

\begin{document}

\maketitle
\noindent \textbf{Preprint version. Submitted to Publications of the Astronomical Society of Australia (PASA).}

\begin{abstract}
This work introduces \ShortName, a proof-of-concept Machine Learning model, leveraging Transformer Encoder architecture to efficiently determine the stellar age, metallicity, and dust attenuation of galaxies from optical spectra. Designed to address the challenges posed by the vast datasets produced by modern astronomical surveys, \ShortName~offers a significant improvement in processing speed while maintaining accuracy. Using the E-MILES spectral library, we generate a dataset of 111,936 diverse templates by expanding the original 636 simple stellar population models with varying extinction levels, combinations of multiple spectra, and noise modifications. This ensures robust training over the spectral range of 4750–7100~\AA~at a resolution of 2.5~\AA. \ShortName~architecture employs four parallel attention-based encoders with varying kernel sizes to capture diverse spectral features. The model demonstrates a mean squared error (MSE) of 0.27\% with a standard deviation of 0.10\% between the input spectra and the \ShortName-generated spectra for the synthetic test dataset. Performance evaluation against real data from two galaxies in the PHANGS-MUSE survey (NGC4254 and NGC5068) demonstrates its ability to extract physical parameters efficiently, with spectral fit residuals showing a mean of -0.02\% and 0.28\%, and standard deviations of 4.3\% and 5.3\%, respectively. To contextualize these results, we compare age, metallicity and dust attenuation maps generated by \ShortName~with those of pPXF, a state-of-the-art spectral fitting tool. While pPXF achieves robust results, it requires approximately 11 seconds per spectrum. In contrast, \ShortName~processes a spectrum in less than 4 milliseconds—a speedup factor exceeding 2,750, while also consuming 68 times less power per spectrum. The comparison with pPXF maps from PHANGS-MUSE underscores \ShortName's capacity to enhance traditional methods through machine learning, paving the way for faster, more energy-efficient, and more comprehensive analyses of galactic properties. This study demonstrates the potential of \ShortName~as an efficient, scalable, and sustainable solution for processing large astronomical surveys.

\end{abstract}


\section{\textbf{Introduction}}

\subsection{\textbf{Background and Motivation}}

Understanding the star formation history (SFH) of galaxies is fundamental to unravelling the processes that shape their formation and evolution. Reconstructing a galaxy's SFH enables us to trace when its stars formed, how its chemical composition evolved, and how various internal mechanisms and external influences drove its growth. Such insights are crucial for addressing broader questions in astrophysics, from galaxy formation and assembly to the role of environment and feedback processes in shaping cosmic structures.


Over the past few decades, significant advancements have been made in the reconstruction of SFHs for both nearby and distant galaxies. This progress has been driven not only by an increase in the quality and quantity of observational data but also by improvements in numerical techniques used to compare these data with stellar population synthesis models. Surveys across a wide range of redshifts have provided an unprecedented view of galaxies' star formation and chemical enrichment histories. However, challenges remain, particularly due to intrinsic spectral degeneracies that make it difficult to disentangle the effects of stellar ages, metallicities, and dust attenuation. These complexities, combined with observational uncertainties, continue to pose significant hurdles to accurately reconstructing SFHs.

A promising breakthrough in this area has come with the advent of high-resolution integral field spectroscopy (IFS). IFS surveys, such as PHANGS (Physics at High Angular Resolution in Nearby Galaxies) \citet{2022A&A...659A.191E} have revolutionised our ability to study nearby galaxies in exceptional detail, resolving structures down to 100-pc scales. These observations provide spatially resolved spectra, capturing both stellar populations and interstellar medium properties across entire galaxies. While this leap in resolution and detail offers the potential to overcome previous limitations, it also introduces new challenges - particularly in managing the vast amounts of data and the complexity and diversity of the spectral features that must be modelled. This is even more important now that the number of galaxies for which such data will become available is rapidly increasing with the advent of surveys such as MAUVE \citep{2024MNRAS.530.1968W} and GECKOS \citep{van2022geckos}.

Traditional techniques for extracting galaxy parameters, such as least-squares fitting or Bayesian minimisation, such as BAGPIPES \citep{carnall2018}, pPXF \citep{cappellari2004parametric},  have played a crucial role in this field. These methods involve fitting observed spectra with theoretical or empirical models to derive key parameters, such as stellar age and metallicity. However, they are often computationally intensive and difficult to scale to the large datasets now being generated by modern surveys. This computational burden and the time-intensive nature of these methods limit their applicability to larger and more complex datasets.

Excitingly, the advent of machine learning (ML) represents a promising avenue to solve some of the technical challenges faced by this field. 
Recent advances in ML have positioned it as a powerful alternative to conventional techniques. ML models, particularly Convolutional Neural Networks (CNNs), have shown significant promise in astronomical applications such as galaxy classification \citep{pasquet2019} and redshift estimation \citep{lovell2019}. In the domain of spectral analysis, approaches like STARNET \citep{fabbro2018starnet} have demonstrated their ability to outperform traditional methods, offering faster and more accurate parameter estimation while reducing computational overhead. Additionally, ML models minimize manual intervention, making them highly suitable for automating the analysis of next-generation astronomical surveys.

In this paper, we present the first, proof-of-concept, realisation of GalProTE, a Machine Learning model leveraging Transformer Encoder architecture to extract stellar population information from integral-field-spectroscopic data. The primary driver for GalProTE is to develop an efficient and reliable tool to analyse resolved observations of nearby galaxies and simultaneously combine input from different stellar population models to encapsulate the diversity of galactic properties, and better understand biases and limitation of current, most standard, approaches. As such, this paper provide a first high-level introduction to the general philosophy and architecture of GalProTE and an example of first application to real observations. 

Indeed, to validate the results of our proposed model, we utilize pPXF, a state-of-the-art and widely adopted method known for its robustness in spectral fitting. By comparing with this well-established approach, we ensure a solid foundation for consistent and reliable parameter estimation. This allows us to illustrate how ML can not only complement traditional methods but also surpass them in terms of speed and scalability. Our goal in this work is not to create a speed-up version of pPXF, but to provide a ML based framework that builds on strengths of pPXF (making it possible a fair comparison between the two) while addressing its limitations, particularly in handling the large-scale datasets produced by modern surveys.

\subsection{\textbf{The Overview of Proposed Method: \ShortName}}

Traditional methods for extracting galactic parameters, such as spectral fitting with empirical or theoretical templates, are computationally intensive and time-consuming, particularly when applied to large-scale surveys. To address this limitation, we propose a deep learning-based approach for estimating age, metallicity, and dust attenuation from galaxy spectra. Our model, \ShortName, is trained on the E-MILES simple stellar population model spectra \citep{vazdekis2016uv}, which includes a comprehensive range of stellar ages and metallicities. We aim to train and test the ML model with a diverse and realistic dataset, by generating combination templates from E-MILES spectra with varying dust attenuation and noise levels.

\begin{figure}[t!]
    \centering
    \includegraphics[width=0.35\textwidth]{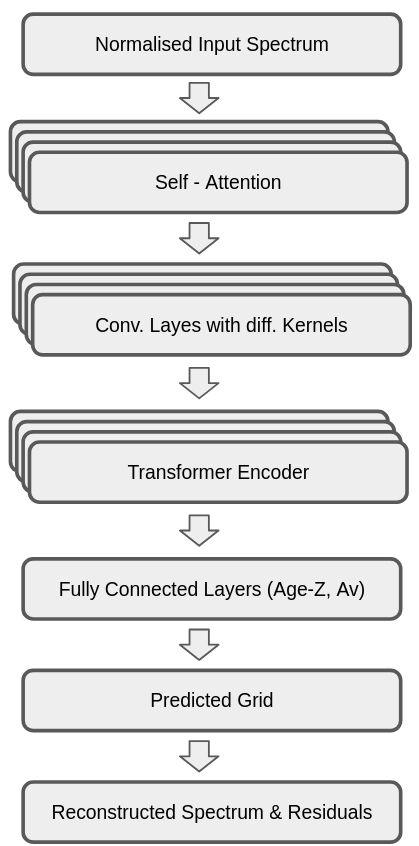}
    
    \caption{Overview of the proposed deep learning model. \ShortName~accepts a normalised spectrum as input, which is processed through four independent parallel blocks. Each block employs a self-attention mechanism to emphasize distinct regions of the spectrum and convolutional layers to extract features. The transformer within each block encodes these features, which are then passed to two fully connected layers. One layer predicts the age-metallicity grid for the input spectrum, while the other predicts the dust attenuation. Using these predictions, a reconstructed spectrum is generated for comparison with the input spectrum.}
    \label{fig:BlockDiagram}
\end{figure}

The synthetic dataset used to train and test the model contains 111,936 spectral templates, spanning ages from 30.0 MYr to 14.0 GYr and metallicities, [M/H] from -2.27 to 0.4, both sampled logarithmically, covering the wavelength range from 4750~\AA~to 7100~\AA~with a resolution of 2.48~\AA. Dust attenuation ranges from $A_v =$ 0 to 1.5 magnitudes, in steps of 0.1, using the attenuation curve from \citep{o1994rnu}. The constraints on the chosen age and wavelength ranges are determined by the limitations of the data and the scope of this work. Specifically, the lower age limit of 30.0 MYr reflects the inherent restriction of the E-MILES synthetic stellar population models used for training. Similarly, the lower and upper wavelength limits of 4750~\AA~and 7100~\AA~respectively are dictated by the spectral coverage of the PHANGS-MUSE survey \citep{2022A&A...659A.191E}, which serves as the observational benchmark for this study.  The model architecture, illustrated in Figure \ref{fig:BlockDiagram}, leverages parallel attention-based transformer encoders \citep{vaswani2017attention} with varying kernel sizes to capture spectral features at multiple scales, improving predictions of stellar population properties. It outputs a fine $12 \times 53$ grid of metallicity and age predictions (from here on referred as predicted grid), along with a dust attenuation value for each input spectrum.


To evaluate the model's performance, we tested it on a synthetic dataset and two face-on galaxies from the PHANGS-MUSE survey: NGC4254 and NGC5068. The choice and number of galaxies were not critical, as the primary goal was to assess the model's ability to generate comparable results efficiently. While these two galaxies exhibit sufficient variation in age, metallicity, dust attenuation, SNR, and velocity gradients, any galaxies from the PHANGS-MUSE survey could have been selected for this purpose. While testing these two galaxies, the model demonstrated remarkable computational efficiency, achieving a speedup of over 2,750 times compared to pPXF. Moreover, the comparison of the resulting maps revealed strong consistency between the \ShortName-generated maps and the pPXF-generated maps, validating the model's accuracy.

\section{\textbf{Dataset Preparation}}
\label{Dataset Preparation}
This work utilises the Medium-resolution Isaac Newton Telescope Library of Empirical Spectra (E-MILES) as the training dataset. The E-MILES dataset is a comprehensive library of empirical stellar spectra covering a broad range of stellar parameters \citep{sanchez2006medium}. It has been extensively used in stellar population synthesis to create models that simulate the integrated light of galaxies. Previous studies have focused on deriving the fundamental properties of stellar populations in galaxies, such as age, metallicity, and initial mass function \citep{vazdekis2010miles, falcon2011miles}. These studies have demonstrated the utility of the MILES/E-MILES library in improving the accuracy of spectral fitting and population synthesis models. 

In the E-MILES dataset, each spectrum/template represents a simple stellar population at a given age and metallicity. The proposed method utilises BaSTI (Bag of Stellar Tracks and Isochrones) to model the evolutionary stages of stars. The initial mass function (IMF) used in this study is the Chabrier IMF (`ch') \citep{chabrier2001galactic}, which describes the distribution of stellar masses at the time of their formation.

The E-MILES dataset covers a range of 12 metallicities and 53 ages, resulting in a total of 636 $(12 \times 53)$ templates, which provide a fine resolution for modelling purposes. These 636 original templates are used as a base to generate a much bigger dataset. The specific log of metallicity [M/H], (from here on Z) and age values in GYr in the E-MILES dataset are:

Z = [-2.27, -1.79, -1.49, -1.26, -0.96, -0.66, -0.35, -0.25, 0.06, 0.15, 0.26, 0.4]

Ages = [0.03, 0.04, 0.05, 0.06, 0.07, 0.08, 0.09, 0.1, 0.15, 0.2, 0.25, 0.3, 0.35,  
0.4, 0.45, 0.5, 0.6, 0.7, 0.8, 0.9, 1.0, 1.25, 1.5, 1.75, 2.0, 2.25, 2.5, 2.75, 3.0, 3.25, 
3.5, 3.75, 4.0, 4.5, 5.0, 5.5, 6.0, 6.5, 7.0, 7.5, 8.0, 8.5, 9.0, 9.5, 10.0, 10.5, 11.0, 
11.5, 12.0, 12.5, 13.0, 13.5, 14.0]

Our immediate focus is a detailed comparison with, and future application to MUSE observations, hence performance of \ShortName~is evaluated on the data of galaxies from PHANGS-MUSE survey. To correct for the instrumental effects of the MUSE on the E-MILES templates, we assume the Line Spread Function (LSF) using Equation 8 from \citep{2017A&A...608A...1B}.

\begin{equation}
\text{LSF}_{\text{MUSE}}(\lambda) = 5.866 \times 10^{-8} \lambda^2 - 9.187 \times 10^{-4} \lambda + 6.040
\end{equation}

The difference between LSF of MUSE, and E-MILES (with LSF of 2.5) was used to calculate Sigmas of Gaussian filters to be applied.

\begin{equation}
\sigma(\lambda) = \left( \frac{\sqrt{\max(\text{LSF}_{\text{MUSE}}(\lambda)^2 - \text{LSF}_{\text{E-MILES}}(\lambda)^2, 0)}}{2.355} \right)
\end{equation}

These filters were applied to the templates, which were then interpolated onto a standardised wavelength range from 4749.37~\AA~to 7100.49~\AA, with a sampling interval of 2.48~\AA, resulting in 948 samples per spectrum.

\subsection{\textbf{Addition of Extinction}}
Each spectrum within the E-MILES dataset represents a unique stellar population with a single age and metallicity, and assumes zero dust extinction. To accurately model the observed spectral characteristics of galaxies, we need to account for the effects introduced by interstellar dust. We selected a range of dust attenuation values $A_v$ from 0.0 to 1.5 in 16 increments of 0.1, based on the extinction law proposed by \citep{o1994rnu} with an assumed $R_v=3.1$. This range aligns with the typical observational conditions for moderately inclined nearby galaxies. By incorporating these 16 extinction levels, we generated 636 single-population templates per extinction level. This ensures robust coverage across a range of extinction scenarios encountered in most astronomical observations. For more extreme scenarios, such as edge-on galaxies or higher-redshift dusty systems, higher $A_v$ values might be necessary. Such conditions fall outside the range for which \ShortName~is trained and validated, and could result in abnormal features in the machine learning results. This will be subject to future work to expand our approach and make it applicable to a wider range of objects.

\subsection{\textbf{Generation of Combination Templates}}

After incorporating various extinction levels, we generated combination templates of multiple populations to better represent the spectral diversity observed in real galactic data. However, the underlying assumption in generating these combinations is that diverse stellar populations represented within a single template share a common extinction level. This is, of course, simplistic but the same approach is used by pPXF, allowing a more consistent comparison. Thus, templates corresponding to the same extinction value were combined in varying numbers, from 1 (single population) up to 5 templates/populations per extinction value. By supporting up to five population combinations per template, we achieve a balance between capturing spectral complexity and managing computational demands. Each combination consisted of randomly selected templates from the \textbf{636} original templates belonging to the same extinction level, with random weights assigned to each template. The sum of these weights was normalised to 1. Each weight represented the relative contribution of the respective template to the composite spectrum, ensuring a realistic representation of the spectral mixtures observed in complex stellar populations.

Using the 636 E-MILES templates, a total of 404,496 combinations can be generated when considering just two stellar populations with one set of weights. When weight variations are included, the number of possible combinations becomes theoretically infinite. However, the objective was not to explore every possible combination but to train \ShortName~to effectively recognize and decompose multiple stellar populations in real spectra. To strike a balance between diversity and computational feasibility, total templates per combination set were limited to 1,272, which is twice the number of original templates per extinction level (636 $\times$ 2). This selection ensures a sufficiently diverse dataset for training while keeping computational and time demands manageable. As more powerful GPUs becomes available for training, these combination templates can be increased to generate a larger synthetic dataset and further improve the model's training potential.

Each original template was replicated two additional times, resulting in three original copies per extinction level. This was done to ensure \ShortName~learns features in their purest form. Random noise added before training (section \ref{Noise}) can obscure features, but replication reduces the chance that noise will affect the same features in all three copies, preserving feature integrity. 

The process of combining templates was randomized, including both the selection of templates and their corresponding weights, to ensure \ShortName~was exposed to a broad range of spectral features. This approach enabled the model to learn how to decompose spectra into distinct stellar populations, forming a robust foundation for advanced spectral analysis of galaxies. Table \ref{tab:template_summary} summarizes the composition of Original (plus two replications) and combined templates for each extinction level, resulting in 6,996 templates per level across 16 extinction levels. The final dataset, consisting of 111,936 templates (6,996 $\times$ 16) along with their corresponding weight grids, was used as a training, validation, and test dataset for \ShortName.

\begin{table}[h!]
    \centering
    \footnotesize 
    \caption{Breakdown of Single and Combination Templates for each value of $A_v$}
    \begin{tabular}{|l|r|}
        \hline
        \textbf{Description} & \textbf{Number of Templates} \\
        \hline
        Original templates per extinction & 636 \\
        \hline
        Replication of original templates (2 times) & 636 $\times$ 2 = 1272 \\
        \hline
        Combinations of 2 random populations & 636 $\times$ 2 = 1272 \\
        \hline
        Combinations of 3 random populations & 636 $\times$ 2 = 1272 \\
        \hline
        Combinations of 4 random populations & 636 $\times$ 2 = 1272 \\
        \hline
        Combinations of 5 random populations & 636 $\times$ 2 = 1272 \\
        \hline
        \hline
        Total Templates per extinction level & 6996 \\
        \hline
    \end{tabular}
    \label{tab:template_summary}
\end{table}

\section{\textbf{Methodology}}
\label{Methodology}
\subsection{\textbf{Processing of Grids}}
\label{Data Pre-processing}

The model shown in Figure \ref{fig:BlockDiagram}, is designed to predict an age-metallicity grid and an $A_v$ value for the input spectrum. The training set consists of anywhere between 1 to 5 populations per template, resulting in 1 to 5 non-zero weights on a $12 \times 53$ grid. However, grids derived from actual galactic spectra are rarely this sparse, as even single populations appear as tight clusters of neighbouring populations on the age-metallicity grid. To enhance the realism of the generated population grids and better reflect the clustering observed in actual galactic data, we implemented a two-step process to generate the final training grids.

First, the weights drawn for the grid were scaled to incorporate realistic priors based on observations of nearby galaxies \citep{neumann2021sdss}. Specifically, weights across metallicity values were linearly scaled such that the lowest metallicity value of $-2.27$ was scaled by a factor of 0.7, while the highest metallicity value of $0.4$ was scaled by a factor of 1.4. Linear interpolation was applied to assign scaling factors to intermediate metallicity values, ensuring a smooth transition across the grid. This adjustment introduced a bias toward higher metallicity values, which are more prevalent in galaxies like those in the PHANGS-MUSE sample, while maintaining representation of lower-metallicity solutions. The adjusted weights were then normalized to ensure that the total weight of the grid was equal to one.

\begin{figure}[t!]
\centering
\begin{minipage}{0.70\linewidth}
\centering
\includegraphics[width=\linewidth]{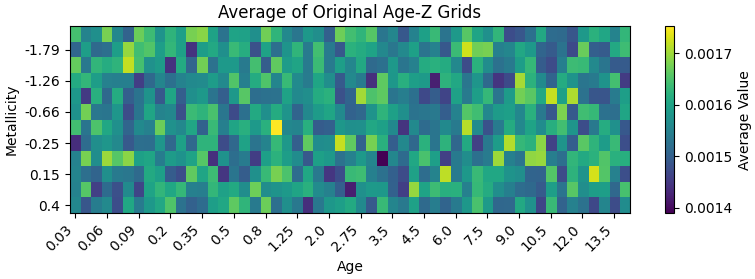}
\end{minipage}\\[1ex] 
\begin{minipage}{0.70\linewidth}
\centering
\includegraphics[width=\linewidth]{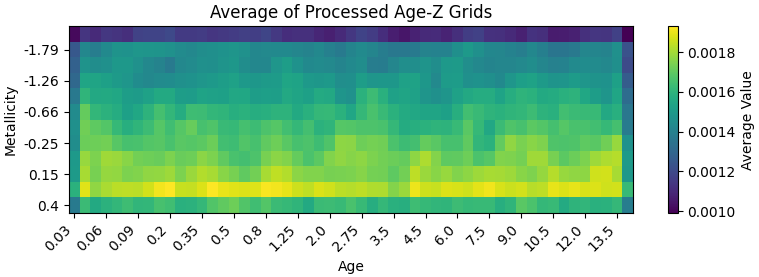}
\end{minipage}

\caption{Average of Age-Z grids for the entire dataset, before and after the processing.}

\label{fig:grids}
\end{figure}

Second, the normalized grids were smoothed to reflect the clustering of populations typically observed in real galactic data. A Gaussian filter with dimensions of $3 \times 3$ and a sigma value of 0.7 bins was applied to the grids. This filter transformed individual grid points into clusters of nine neighbouring populations while preserving the dominance of the original bin in the cluster. The Gaussian kernel was normalized to ensure that the combined weight of the resulting population cluster remained consistent with the original weight. This smoothing process created more realistic distributions of stellar populations while retaining the priors introduced in the first step.

Figure \ref{fig:grids} displays the average of the original and adjusted grids across the entire dataset in the age-metallicity plane. The adjustments shift the prior distribution to emphasize metal-rich solutions, aligning with the observed characteristics of nearby galaxies such as those in the PHANGS-MUSE sample. Importantly, the adjustments do not exclude metal-poor solutions but reduce their relative weight to reflect their lower prevalence in these galaxies. The adjusted grids maintain continuity across age values while appropriately representing more metal-rich solutions, which dominate the sample. This approach ensures that the model incorporates prior knowledge while retaining flexibility to explore the full range of parameter space.

Finally, the template spectra associated with each grid were recalculated based on the adjusted and smoothed grid weights and the associated extinction values $A_v$. This recalculation ensured that the spectra accurately represented the modified distributions of stellar populations.

\subsection{\textbf{Addition of Noise to Spectra}}
\label{Noise}
To enable \ShortName~to recognise spectral features both in their pure form and as affected by varying levels of noise, we introduced noise to the spectra in the dataset in a structured manner. Each template (single or combination) spectrum was first normalised by dividing it by its median flux value. Noise was then added to the template spectra, while the respective grids were kept unchanged. This approach was justified by the need to simulate realistic observational conditions. Typically, spectra are binned together to maximise the signal-to-noise ratio (SNR) and bring it to a level of 30 to 35 \citep{pessa2023resolved}, corresponding to a noise level of approximately 3.3\%. Additionally, there are residuals from skylines subtraction, such as the oxygen line at 6300~\AA, which are not included in E-MILES template spectra. By incorporating noise and simulating the effects of median filtering of bright sky subtraction, we aimed to replicate real observational conditions, allowing the model to learn to handle and account for these variations.

For the first phase of noise addition, the entire dataset was randomly shuffled and divided into three equal sets to be treated differently:

\begin{itemize}
\item The first set had no noise added to preserve features.
\item The second set was added with a Gaussian noise of 2\%.
\item The third set had 4\% Gaussian noise added to it.
\end{itemize}

To prevent \ShortName~from learning to rely solely on specific and prominent features during training, a second phase of noise was added with an aim to significantly alter random features. This noise addition involved two types of modifications to the spectra:

\begin{itemize}
\item \textbf{High Noise Patches:} Randomly selected patches of spectra were subjected to Gaussian noise of up to 25\%, simulating the presence of residuals of skylines subtraction, or other filters applied during pre-processing. 
\item \textbf{Masked Patches:} Randomly selected patches of spectra were replaced with the median value of the respective patch, imitating the masked data often encountered in real observations.
\end{itemize}

The total number of random patches added to each spectrum ranged from 15 to 25. Each patch was randomly decided to be either a high noise patch or a masked patch. The location and width of each patch were also randomly chosen, varying between 3 and 15 spectral bins, which corresponds to a width of 7.44~\AA~to 37.2~\AA. For noisy patches, Gaussian noise with a random std deviation between 0 and 25\%, but zero mean was applied to ensure that the noise addition did not introduce any systematic bias into the dataset. This combination of noise addition techniques helped the model to generalise effectively across a diverse range of noisy and incomplete data scenarios.

\begin{table}[h!]
    \centering
    \footnotesize
    \caption{Masked spectral regions to exclude major emission lines.}
    \begin{tabular}{|l|l|c|}
        \hline
        \textbf{Emission Line(s)} & \textbf{Mask Range (\AA)} & \textbf{Width (\AA)} \\
        \hline
        H$\beta$ & 4853.64--4866.06 & 12.42 \\
        \hline
        [O~\textsc{iii}]  & 4952.95--4965.37 & 12.42 \\
        \hline
        [O~\textsc{iii}]  & 5000.12--5012.54 & 12.42 \\
        \hline
        Sky (OH lines) & 5526.46--5553.78 & 27.32 \\
        \hline
        Sky (OH lines) & 6243.96--6256.38 & 12.42 \\
        \hline
        [O~\textsc{i}]  & 6293.60--6306.02 & 12.42 \\
        \hline
        [N~\textsc{ii}], H$\alpha$, [N~\textsc{ii}]  & 6546.85--6589.05 & 42.20 \\
        \hline
        [S~\textsc{ii}] & 6710.70--6738.02 & 27.32 \\
        \hline
    \end{tabular}
    \label{tab:masking}
\end{table}

\subsection{\textbf{Masking of Emission Lines}}
\label{Masking}
The addition of high-noise patches to the training data helps mitigate the impact of residuals from skyline filtering/subtraction, which shift their position based on redshift correction. While the E-MILES templates model stellar populations, real galaxy spectra also contain nebular emission lines that are not included in these templates. Hence, to avoid incorrect results, it is essential to mask these regions of the spectra during the fitting process. Table \ref{tab:masking} describes the regions that have been masked from the spectra. The masking process involves completely removing these bands from the spectra and stitching the remaining parts together to avoid any discontinuities. A total of 72 samples are masked from the original spectra with 948 samples, resulting in a reduced sample size of 876 samples per spectra.

\begin{figure}[h!]
    \centering
    \includegraphics[width=0.60\linewidth]{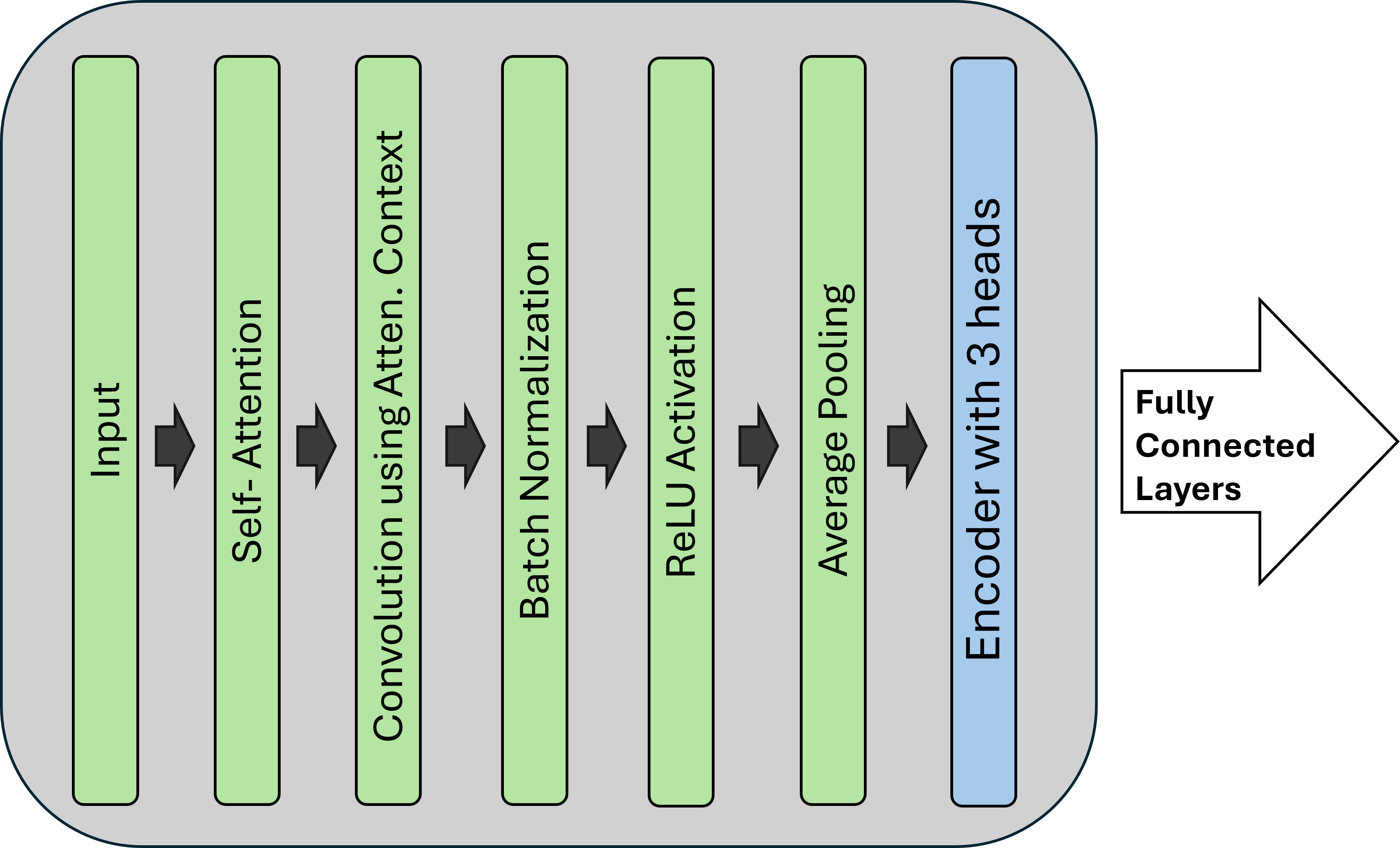}
    \caption{Block diagram of the model's data flow from input to the Encoder's output. The self-attention mechanism extracts context from input spectrum, which is then added back to the input spectrum, followed by convolution, batch normalisation, and ReLU activation. After average pooling, data passes through the Transformer Encoder layer with three attention heads, producing the block output.}
    \label{fig:ModelDiagram}
\end{figure}

\subsection{\textbf{\ShortName~Architecture}}
The model architecture is designed to determine the age-metallicity grid and dust attenuation of galaxies or galactic regions by processing their spectral data. The architecture leverages techniques such as self-attention encoders and multiscale feature extraction to handle the complexity and diversity of galactic spectra. Figure \ref{fig:ModelDiagram} shows the core structure of a single block used in the model. Four such blocks operate in parallel with different kernel sizes, and their outputs are concatenated before being passed to two fully connected layers for Age-Z grid and dust attenuation predictions. The following sections provide a detailed overview of the key components and processing steps involved, highlighting how each part contributes to the overall prediction task.

\subsubsection{\textbf{Self-Attention Mechanism}}

\ShortName~employs a self-attention mechanism to compute context vectors that highlight relevant features in galactic spectra for predicting age, metallicity, and dust attenuation parameters. The self-attention mechanism enables the model to weigh different parts of the input spectrum and capture important patterns at various wavelengths, which are crucial for the prediction of astrophysical parameters.

The computation for context vectors proceeds as follows:

\begin{itemize} 

\item \textbf{Projections}: The input galactic spectrum is passed through linear layers, which learn to project the spectrum into three different representations: Query (Q), Key (K), and Value (V). These projections are learned parameters, meaning that during training, the model adjusts these linear layers to capture relevant spectral features that are important for distinguishing between different parameters.

\item \textbf{Normalisation}: After the projections, each of the Q, K, and V representations undergoes layer normalisation. This helps stabilise the learning process by ensuring consistent feature distributions across batches of spectra.

\item \textbf{Attention Scores}: The model computes attention scores by taking the dot product between the Query (Q) and Transpose of Key ${(K^T)}$ vectors. This measures the similarity between different parts of the spectrum, helping the model identify relationships between certain wavelength regions. These scores are scaled by the square root of the hidden dimension ${\sqrt{d_K}}$ to avoid excessively large values.

\item \textbf{Attention Weights}: The attention scores are passed through a softmax function to produce attention weights. These weights reflect the importance of each part of the input spectrum. The higher the weight for a particular wavelength region, the more influence that region has on the output of the respective block.

\item \textbf{Context}: The attention weights are used to compute context vectors by applying them to the corresponding Value representations. This operation allows the model to focus on the most relevant parts of the input spectrum. These learned Context Vectors effectively summarize this relevant information, which is later used by the model.

The calculations explained above can be formalized for the 4 blocks as
\begin{equation}
\label{eq:attention_computation}
\text{Context}_j = \text{softmax}\left(\frac{Q_j K_j^T}{\sqrt{d_K}}\right) V_j
\end{equation}

Where:

\begin{itemize}
    \item \( j \) represents the block number of \ShortName.
    \item \( d_K \) is the dimension of the Key vector (876, same as length of input spectrum).
    \item \({Context}_j \) is the learned context vector for block j.
\end{itemize}

\end{itemize}

The projections Q, K, and V represent learned embeddings of the input spectrum, enabling the model to identify key spectral features associated with various stellar population parameters. Through backpropagation in the training process, these projections are iteratively refined to improve the accuracy of parameter predictions. For instance, certain spectral regions may have a stronger influence on age estimation, while others may be more indicative of metallicity. The overall spectral shape is particularly important for determining dust attenuation. The extent of these regions can vary, and the self-attention mechanism within each block allows the model to dynamically focus on specific regions/features (with different size) of the spectrum, assigning higher attention weights to areas most relevant for accurate prediction.

\begin{figure}[h!]
    \centering
    \includegraphics[width=0.90\linewidth]{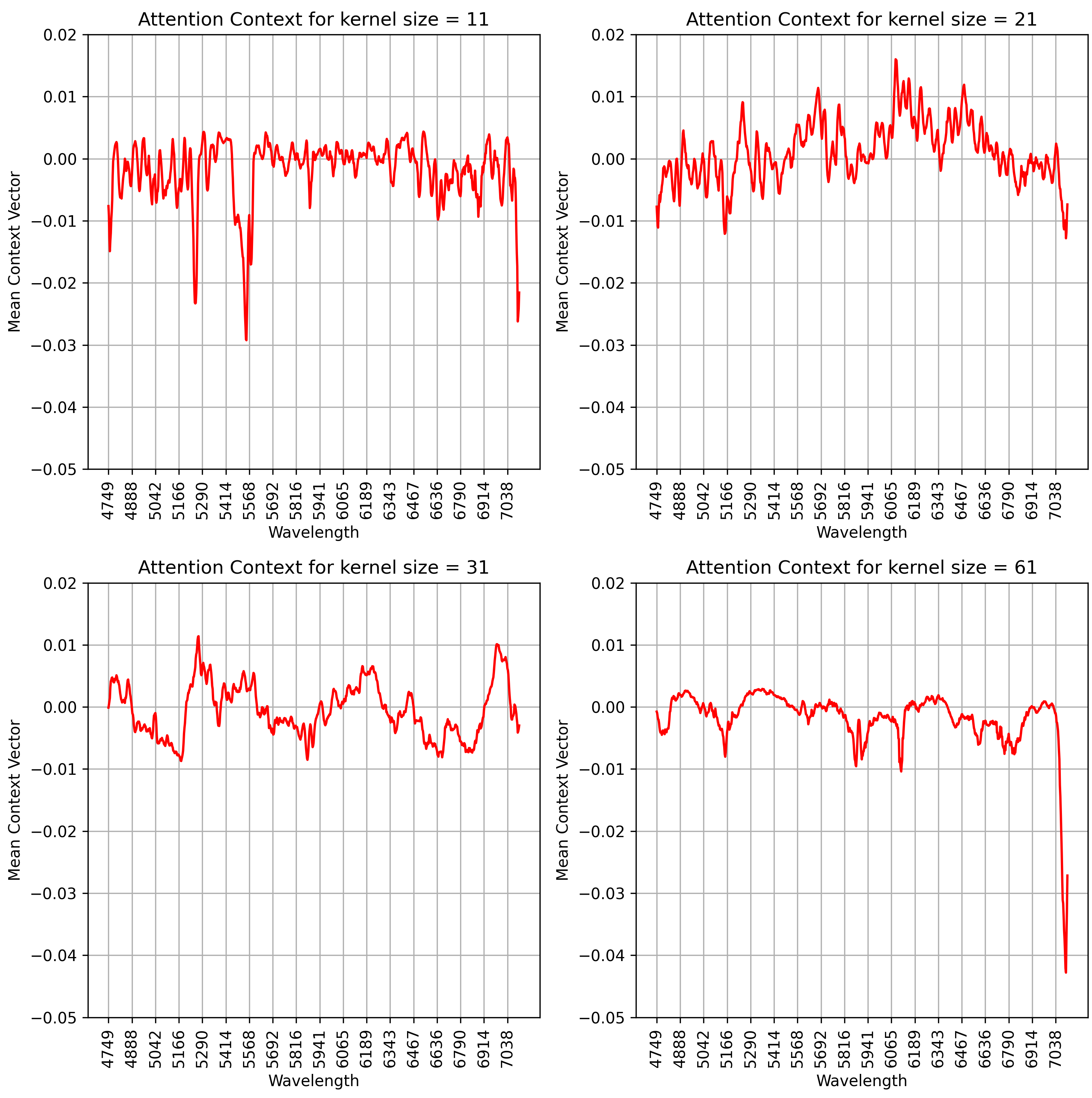}
    \caption{Mean context vectors learned by the 4 parallel blocks of model with different kernel sizes, where each block emphasises features of different sizes. These contexts are added to the input spectra, before features are extracted to predict age, metallicity and dust attenuation.}
    \label{fig:attention_context}
\end{figure}

Figure \ref{fig:attention_context} shows the average context vector learned by the four blocks of \ShortName~for the test dataset. The kernel size in each block determines the scale of features learned. For the kernel size of 11, the model captures high-frequency details, enhancing small, intricate features in the spectra. In contrast, the kernel size of 61 applies a smoother context, focusing on the overall spectral shape. By adding the respective context vector to the input spectra in each block, the model highlights distinct characteristics, which are then processed to extract features.

\subsubsection{\textbf{Parallel Processing using Different Kernels}}

The Context Vector learned by the attention mechanism (Eq. 3) of each block is added to the spectra for further processing. These processing steps comprise of convolutional layers for feature extraction, batch normalisation for stabilisation, rectified linear unit (ReLU) activation for non-linearity, and average pooling to reduce spatial dimensions. Importantly, each block incorporates its own transformer encoder layer that utilises kernel-specific feature extraction based on learned context vector.

\begin{equation}
\text{Context\_Spec}_j = \text{Context}_j + \text{Input\_Spectrum}
\end{equation}

where \( \text{Context}_j \) is the context vector for block \( j \), as computed in Equation~\ref{eq:attention_computation}, and \( \text{Context\_Spec}_j \) is the contextualised spectrum which is fed as input to the Convolution layer for block \( j \).

\begin{equation}
\text{X}_j = \text{ReLU}\left(\text{BatchNorm}\left(\text{Conv}(\text{Context\_Spec}_j, W_j, b_j)\right)\right)
\end{equation}

where \( \text{Conv} \) denotes the convolution operation with kernel size (11, 21, 31, or 61) specific to corresponding block \( j \). And \( W_j \) and \( b_j \) represent the weights and biases of the convolutional layer for block \( j \), respectively.

\begin{equation}
\text{X}_j = \text{AveragePool}(\text{X}_j)
\end{equation}

\begin{equation}
\text{Output}_j = \text{Encoder}\left(\text{X}_j\right)
\end{equation}

where \( \text{Encoder} \) represent the sequential Transformer Encoder applied to \( \text{X}_j\), with 3 heads in the Encoder layer.

The outputs from the four blocks are concatenated and then passed through two separate fully connected layers: one for predicting the metallicity-age grid and another for predicting dust attenuation ($A_v$). While the feature extraction blocks are shared between both tasks, each has a dedicated, fully connected layer for final predictions. The metallicity-age grid encodes the contribution of various ages and metallicities in the spectra, providing a detailed grid of size $12 \times 53$. For dust attenuation, \ShortName~outputs a $1 \times 16$ array representing class probabilities for dust attenuation values between 0.0 and 1.5 (in 0.1 increments), with the value corresponding to the highest probability chosen as the final $A_v$.

\subsubsection{\textbf{Grid Clipping}}
To enhance the accuracy and reliability of the predicted Age-Z grids, we implement a clipping mechanism that removes weights contributing less than 5\% relative to the highest weight in each predicted grid. Noise in the spectra translates into noise on the predicted grid. The clipping of the predicted grid mitigates uncertainties and noise by filtering out the least significant weights. The cutoff threshold can be adjusted to filter out populations from the grid to any desired extent. However, through testing, a 5\% threshold of the maximum weight in the respective grid was identified as the best cutoff based on the MSE of spectral fits. The weights below the 5\% threshold are clipped to zero by the model, reducing the impact of noise in the spectra without suppressing the populations with weights above 5\%. The remaining weights after clipping are normalised to ensure that the sum of weights across each grid equals one. 

\subsection{\textbf{Training, Validation, and Testing}}
The process of training, validation, and testing \ShortName~involves leveraging the dataset with pre-processing steps such as smoothing and weight shifting (section \ref{Data Pre-processing}), addition of noise (section \ref{Noise}), and masking (section \ref{Masking}), with the aim that the model generalises well to real-world galactic spectra. The entire dataset, consisting of 111,936 synthetic spectra with various noise levels, is used for training, validation, and testing. Although the same base spectra are used across these sets, different random noise is added to each set, making each set unique and allowing robust validation and testing.

During the training phase, the model iteratively processes batches of data, updating its parameters and attention weights to minimise the defined loss function. After each epoch, the model's performance is validated using the validation set. To prevent over-fitting, we employ early stopping on grid MSE loss with a patience of 3 epochs. This means that if the model's performance on the validation set does not improve for three consecutive epochs, the training process is halted. This avoids over-training and the model retains its generalisation capability.

Once the training is complete, the final model is evaluated on the test set to assess its performance. The total time required for training, validating, and testing the model, along with generating the evaluation plots, is around 12 and a half hours using an NVIDIA GeForce RTX 4090 GPU with 24 GB of memory, with a power rating of 450W. The results of testing the model are discussed in detail in section 4.

\begin{figure*}[b!]
    \centering
    \includegraphics[width=0.95\textwidth]{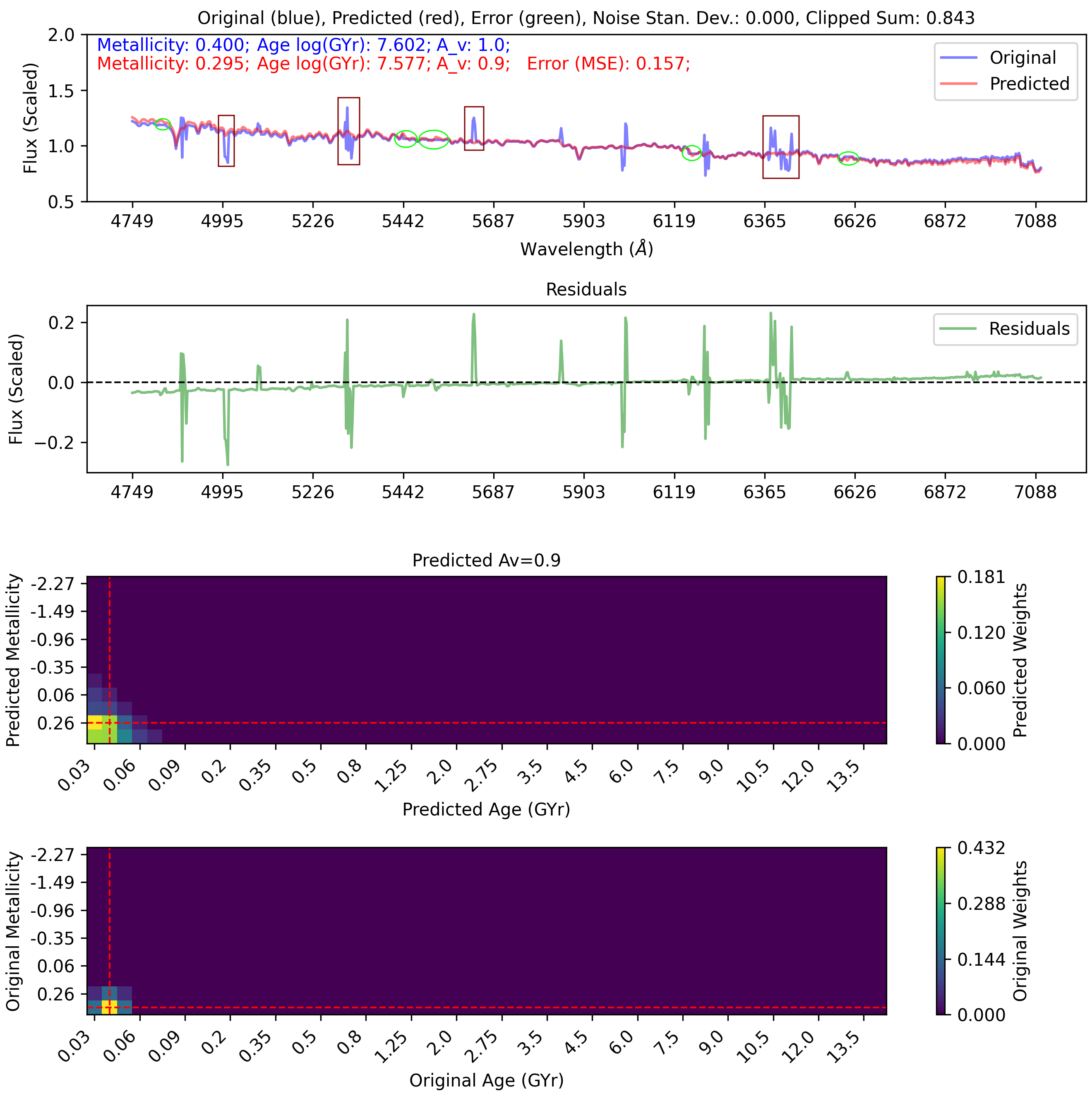}
    \caption{A test spectrum with 0\% Gaussian noise, high-noise patches, and masked patches. The first subplot compares the \textcolor{blue}{noisy input spectrum} with the \textcolor{red}{predicted spectrum}. Some of the high noise patches are shown using brown rectangles, whereas green ellipses show some of the masked patches in the input spectra. The second subplot shows the \textcolor{green}{residuals} between the input and predicted spectra. The third subplot displays the grid predicted by the model, with the mean age and metallicity bin marked. The fourth subplot presents the grid of the original noise-free input spectrum. The residuals are primarily concentrated in high-noise and masked patches, with some impact from the mismatch of 0.1 in Av. Overall, the predicted grid and the reconstructed spectra demonstrate promising accuracy.}
    \label{fig:30500}
\end{figure*}

\begin{figure*}[t!]
    \centering
    \includegraphics[width=0.95\textwidth]{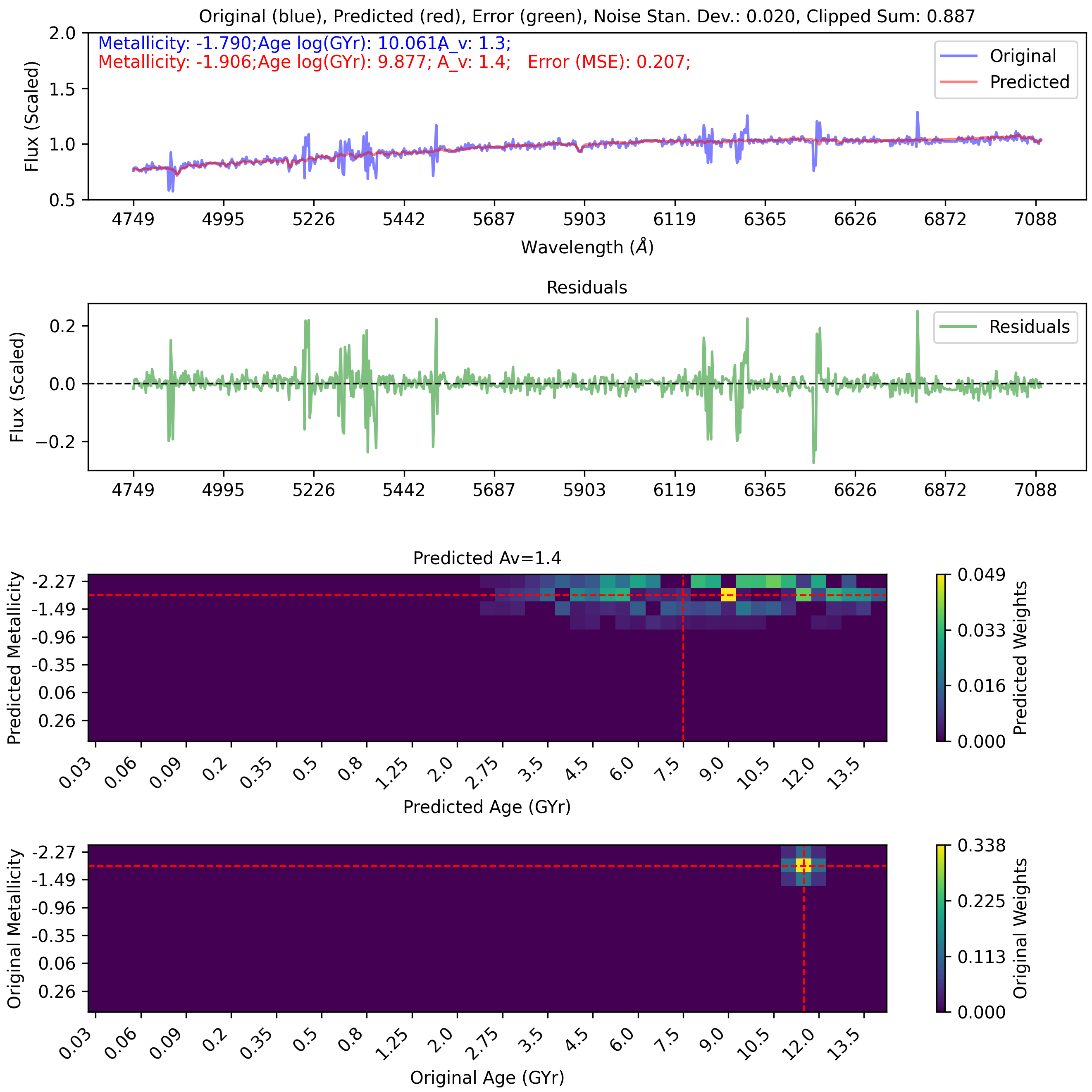}
    \caption{A test spectrum with 2\% Gaussian noise, high-noise patches, and masked patches. The predicted mean metallicity aligns closely with the original metallicity on the grid, while the predicted age shows some error. The model tends to struggle for higher age values. However, the most dominant population on the predicted grid (bright yellow) is notably closer to the original mean age.}
    \label{fig:74250}
\end{figure*}

\begin{figure*}[t!]
    \centering
    \includegraphics[width=0.95\textwidth]{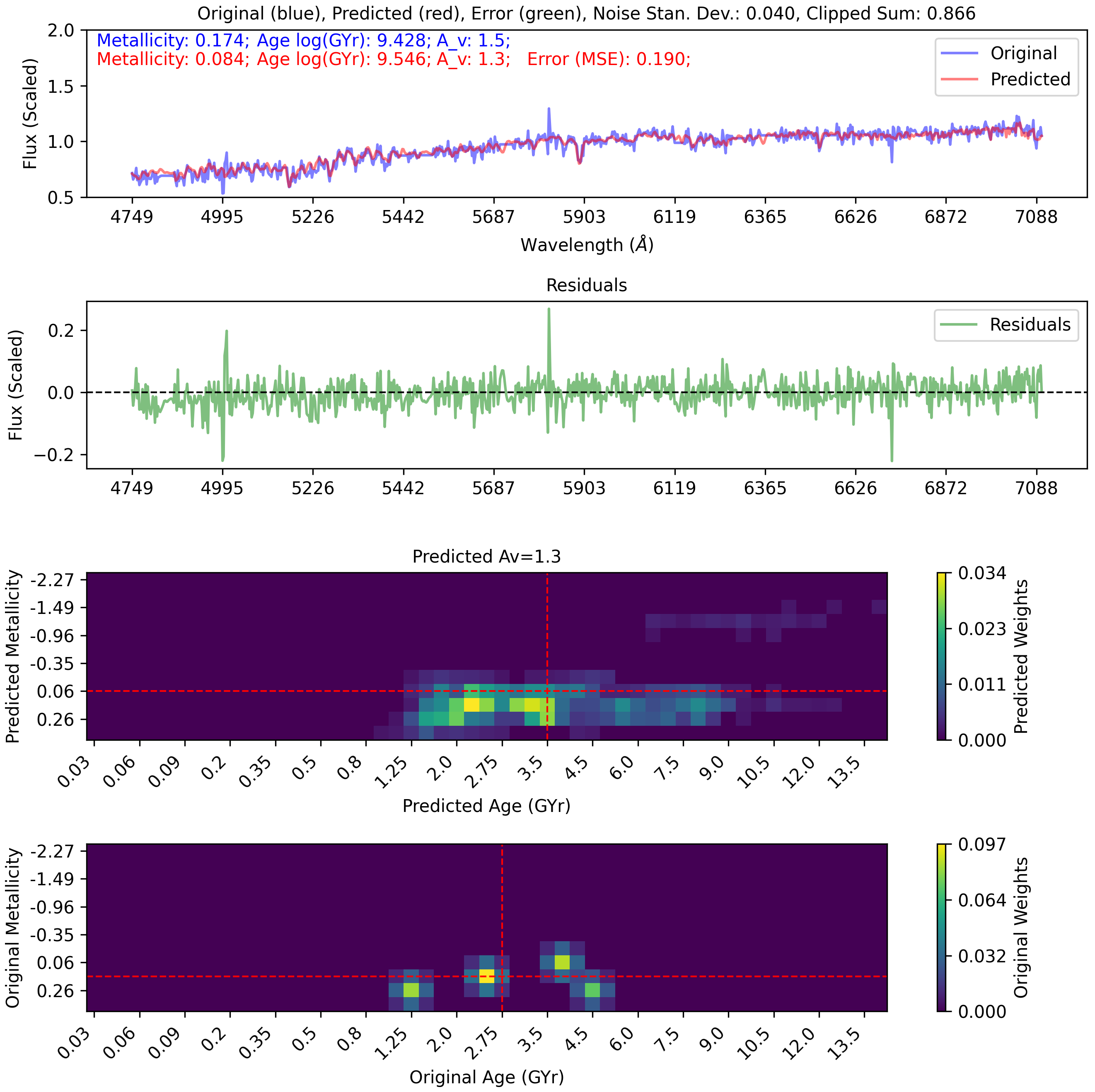}
    \caption{A test spectrum with 4\% Gaussian noise, high-noise patches, and masked patches. This example poses a significant challenge due to the high level of Gaussian noise and the close proximity of the four populations on the original grid. Despite these, the predicted mean values remain close to the original means, and the spectral fit demonstrates reasonable quality, showcasing the model's robustness under adverse conditions.}
    \label{fig:99750}
\end{figure*}

\subsubsection{\textbf{Loss Function}}

The loss function plays a critical role in guiding the model's learning process. For \ShortName, a custom loss function combines Mean Squared Error (MSE) with weighted correlation losses for age and metallicity to direct the model's learning. \textit{MSE Loss} penalises the difference between predicted and target grids, ensuring that the predicted weights on the grid are close to the target weights. \textit{Weighted Correlation Loss} preserves structural correlation within grids by computing \textit{Metallicity Correlation} and \textit{Age Correlation} between the predicted and target grids. Correlation Loss is calculated as $(1 - \text{correlation})$ for both metallicity and age. MSE loss is also employed for the dust attenuation values. This approach, using MSE for individual value accuracy and correlation losses to maintain structural relationships, ensures that the model learns to predict the age-metallicity grids, and dust attenuation values as closely as possible while preserving the inherent correlation.

\section{\textbf{Results and Analysis}}
In this section, we discuss the results of the \ShortName~and analyse the outcomes. Testing \ShortName~with the synthetic E-MILES test set provides valuable insights into the model's training and any potential biases. However, the actual performance of the model can only be assessed when evaluated on real galactic data. As such, in the second part of this section, we will test our model on two galaxies extracted from the PHANGS-MUSE sample: NGC4254 and NGC5068. These face-on galaxies have been chosen to test different ranges of metallicities, star formation rates and noise levels.

This work primarily focuses on fitting the light-weighted spectra to derive stellar ages and metallicities from the predicted grids. The light-weighted spectra are obtained by normalizing the observed spectra using their respective median flux. For the selected range of E-MILES spectra, the median flux is centred at 5886~\AA~for the synthetic data. However, due to the noise in the galactic spectra, the wavelength corresponding to the median flux varies slightly, typically falling within the range of 5876~\AA~to 5896~\AA~for NGC4254 and NGC5068.

\subsection{\textbf{Performance on Synthetic Test Dataset}}
To evaluate the performance of \ShortName, we conducted extensive tests using the synthetic test dataset. This test set consists of the same 111,936 synthetic spectra (section \ref{Dataset Preparation}) which were used for training and validating the model (section \ref{Methodology}), but has its own noise profile (random Gaussian noise of various levels, random high-noise, and masked patches at random locations), which makes it different from the training and validation sets. This subsection presents the results for grid predictions, leading to age, metallicity predictions, as well as the prediction of dust attenuation on the test set. We employ histograms, confusion matrices and MSE distributions to assess the model's effectiveness.

Figures \ref{fig:30500}, \ref{fig:74250}, and \ref{fig:99750} illustrate how \ShortName~performs on test spectra with varying levels of Gaussian noise (0\%, 2\%, and 4\%), as well as high-noise patches and masked regions. The top subplot in each figure shows the \textcolor{blue}{input spectrum} alongside the \textcolor{red}{predicted spectrum}, demonstrating the model's ability to reconstruct the spectra. While the predicted spectra generally align well with the inputs, the \textcolor{green}{residuals} reveal areas of discrepancy, often concentrated in high-noise regions and masked patches. Figure \ref{fig:99750} presents a challenging case of a high noise spectrum, where the 4 populations (shown in the original grid at the bottom) are close to each other. The model struggles to distinguish between these populations in the predicted grid. The input spectrum’s grid suggests a mean metallicity of [M/H] = 0.174 and log(Age) = 9.43, while the predicted grid points to a lower metallicity of [M/H] = 0.084 and a slightly older log(Age) = 9.55. These variations likely result from the model’s attempt to balance the noisy input with the features it learned during training, potentially smoothing out some of the higher metallicity contributions. Similarly, while the original dust attenuation value is 1.5 mag, \ShortName~predicts a slightly lower value of 1.3 mag, possibly due to the model’s sensitivity to noise and its handling of masked regions in the input spectrum. For the mean age and metallicity predictions, the model's performance is very reasonable, even in this challenging case of high noise spectrum and closely spaced populations.

\begin{figure}[t!]
    \centering
    \begin{subfigure}[t]{0.48\linewidth}  
        \centering
        \includegraphics[width=\linewidth]{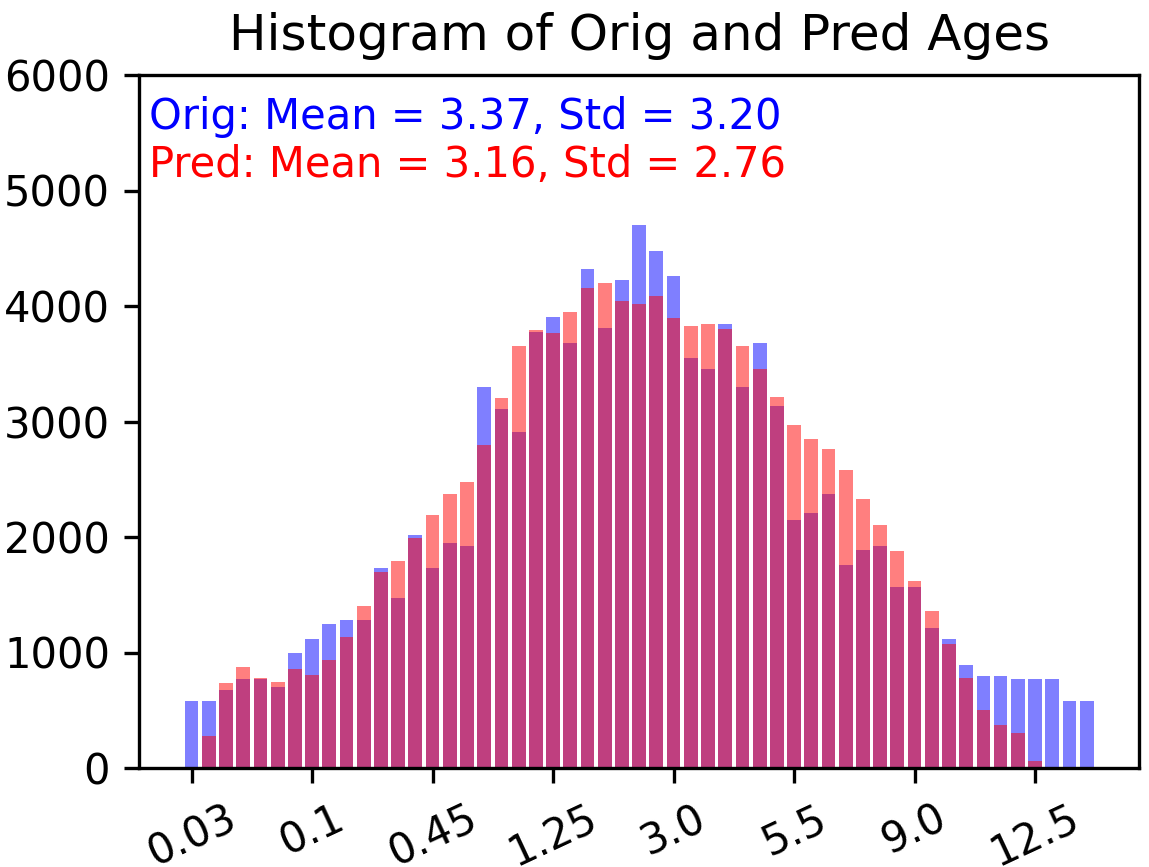}
        \begin{picture}(0,0)
        \put(10,4){\makebox(0,0){\small Age [Gyr]}} 
        \end{picture}
        \begin{picture}(0,0)
        \put(-120,90){\rotatebox{90}{\small Count}} 
        \end{picture}
        \caption{Age distribution histogram.}
        \label{fig:age_histogram}
    \end{subfigure}
    \hfill
    \begin{subfigure}[t]{0.48\linewidth}  
        \centering
        \includegraphics[width=\linewidth]{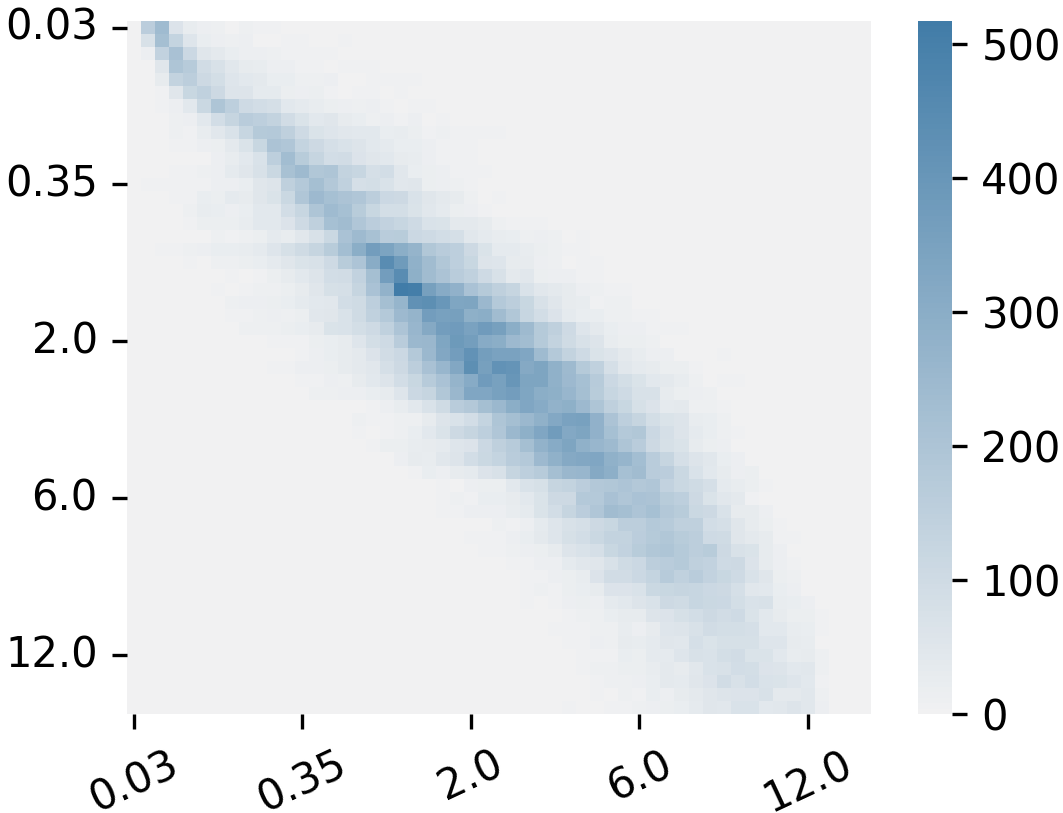}
        \begin{picture}(0,0)
        \put(0,2){\makebox(0,0){\small Age Pred}} 
        \end{picture}
        \begin{picture}(0,0)
        \put(-120,90){\rotatebox{90}{\small Age Orig}} 
        \end{picture}
        \caption{Confusion matrix for predicted vs. original age.}
        \label{fig:cfm_age}
    \end{subfigure}
    \caption{Comparison of the original and predicted mean ages using a histogram (left) and a confusion matrix (right).}
    \label{fig:Test_Age_Error_stats}
\end{figure}
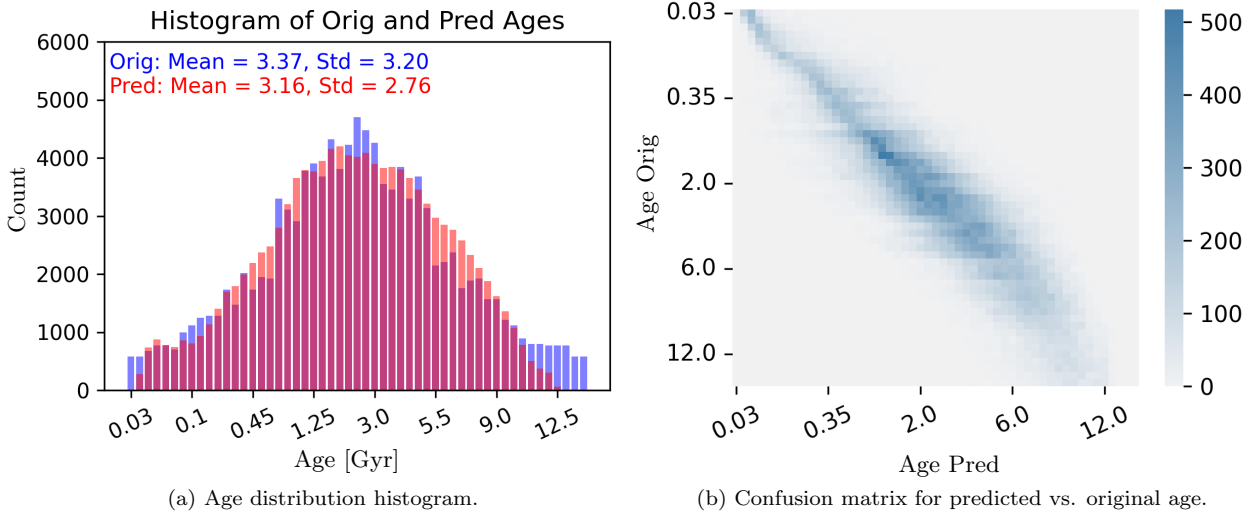

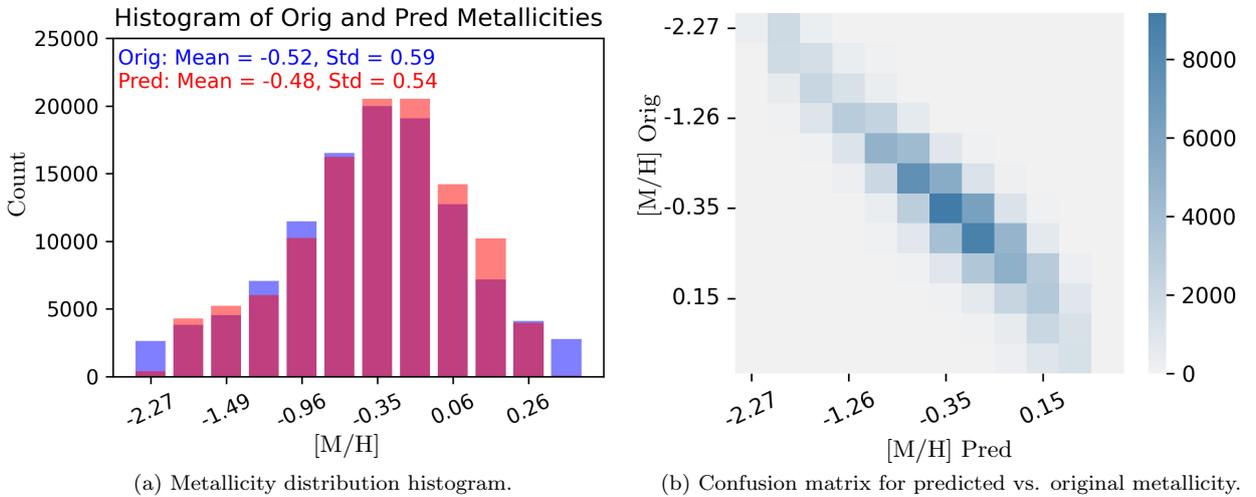
\begin{figure}[b!]
    \centering
    \begin{subfigure}[t]{0.48\linewidth}  
        \centering
        \includegraphics[width=\linewidth]{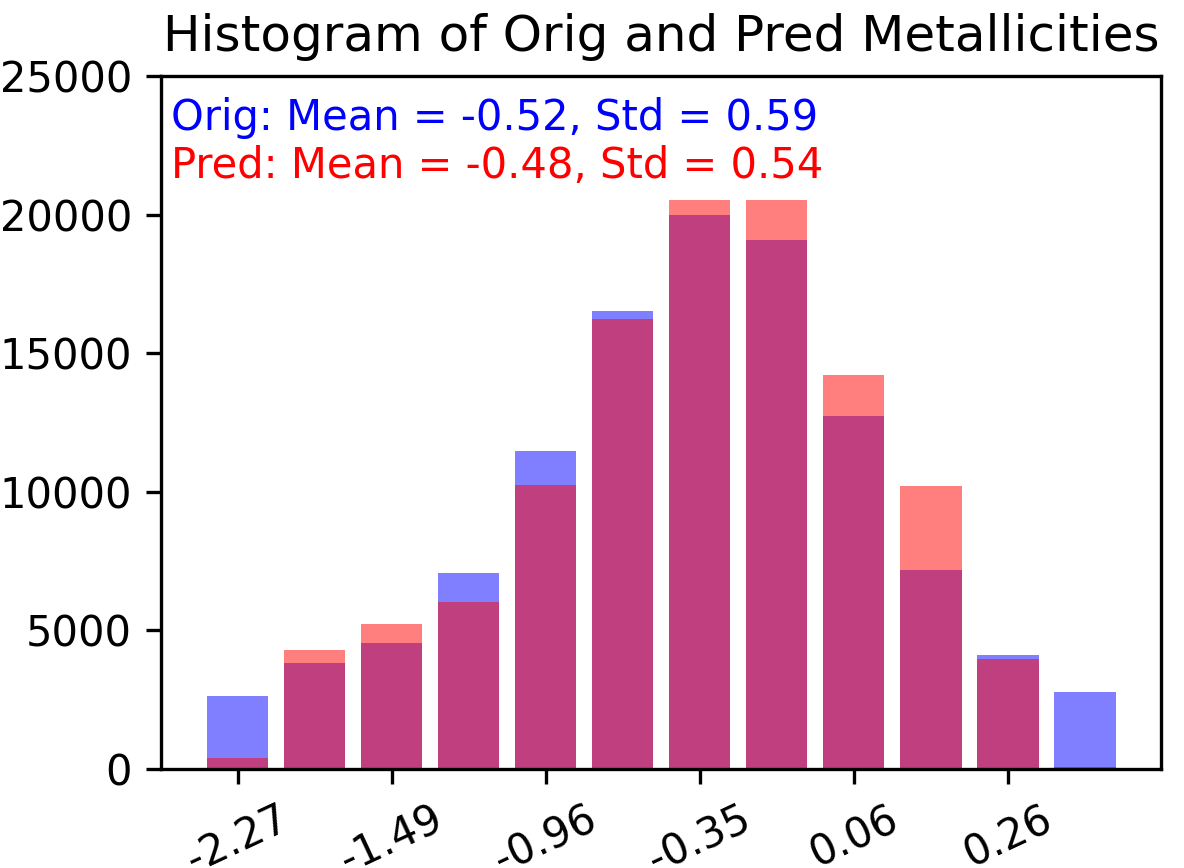}
        \begin{picture}(0,0)
        \put(10,4){\makebox(0,0){\small [M/H]}}
        \end{picture}
        \begin{picture}(0,0)
        \put(-120,90){\rotatebox{90}{\small Count}} 
        \end{picture}
        \caption{Metallicity distribution histogram.}
        \label{fig:Metallicities_histogram}
    \end{subfigure}
    \hfill
    \begin{subfigure}[t]{0.48\linewidth}  
        \centering
        \includegraphics[width=\linewidth]{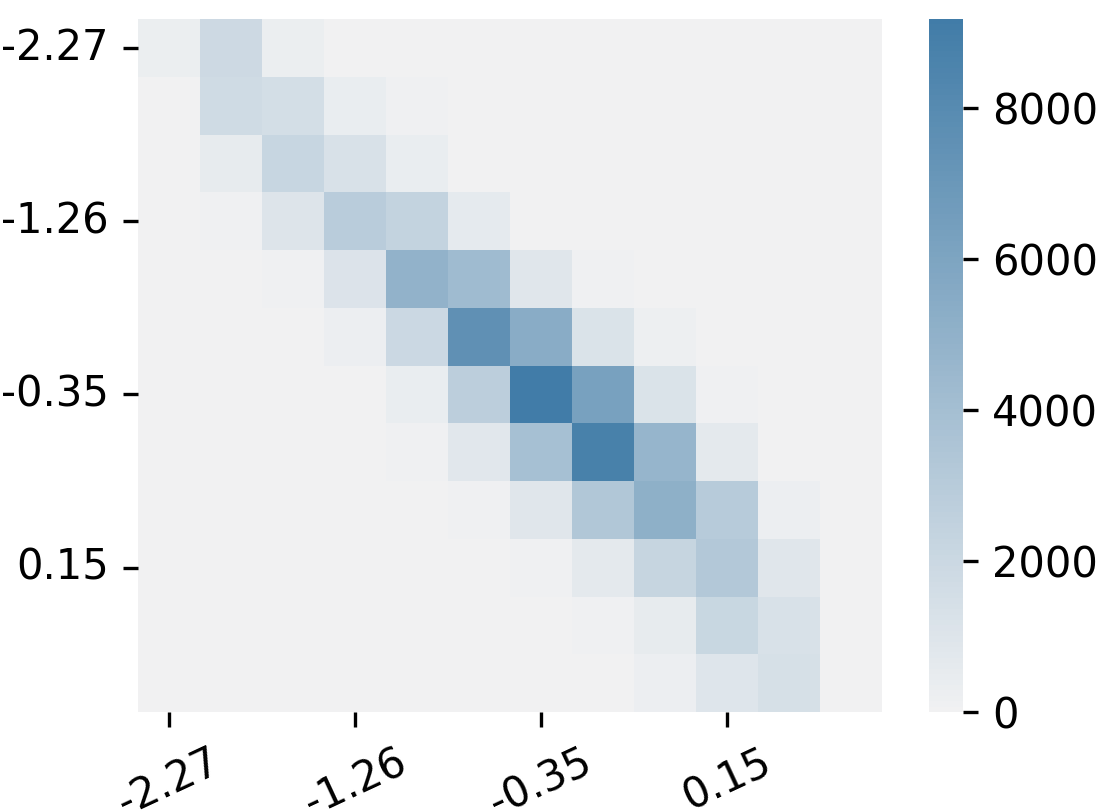}
        \begin{picture}(0,0)
        \put(0,2){\makebox(0,0){\small [M/H] Pred}} 
        \end{picture}
        \begin{picture}(0,0)
        \put(-120,90){\rotatebox{90}{\small [M/H] Orig}} 
        \end{picture}
        \caption{Confusion matrix for predicted vs. original metallicity.}
        \label{fig:cfm_Metallicity}
    \end{subfigure}
    \caption{Comparison of the original and predicted mean metallicities using a histogram (left) and a confusion matrix (right).}
    \label{fig:Test_Z_Error_stats}
\end{figure}

\subsubsection{\textbf{Ages}}

The histogram in Figure \ref{fig:Test_Age_Error_stats} shows the distribution of original and predicted mean ages across the entire test set. For this analysis, the mean age of the populations represented in the templates is used. The overall agreement between the original and predicted age distributions highlights \ShortName's effectiveness in capturing the age characteristics. The difference between the mean age of the original dataset and the predictions is 0.21 GYr. The confusion matrix for predictions has a prominent diagonal dominance, indicating that the model accurately predicts most ages, with minimal confusion between different age bins. However, the model shows limited performance for the age bins on the higher end in the test set, consistently underestimating these values. Conversely, there are overestimates for the lowest age bin of 0.03 GYr. These misclassifications are largely attributed to the training set imbalance, where the mean ages of templates follow a Gaussian-like distribution. Because of this imbalance, the model tends to be biased towards predicting ages towards the centre of this distribution.

\subsubsection{\textbf{Metallicities}}

\begin{figure}[t!]
    \centering
    \begin{subfigure}[t]{0.48\linewidth}  
        \centering
        \includegraphics[width=\linewidth]{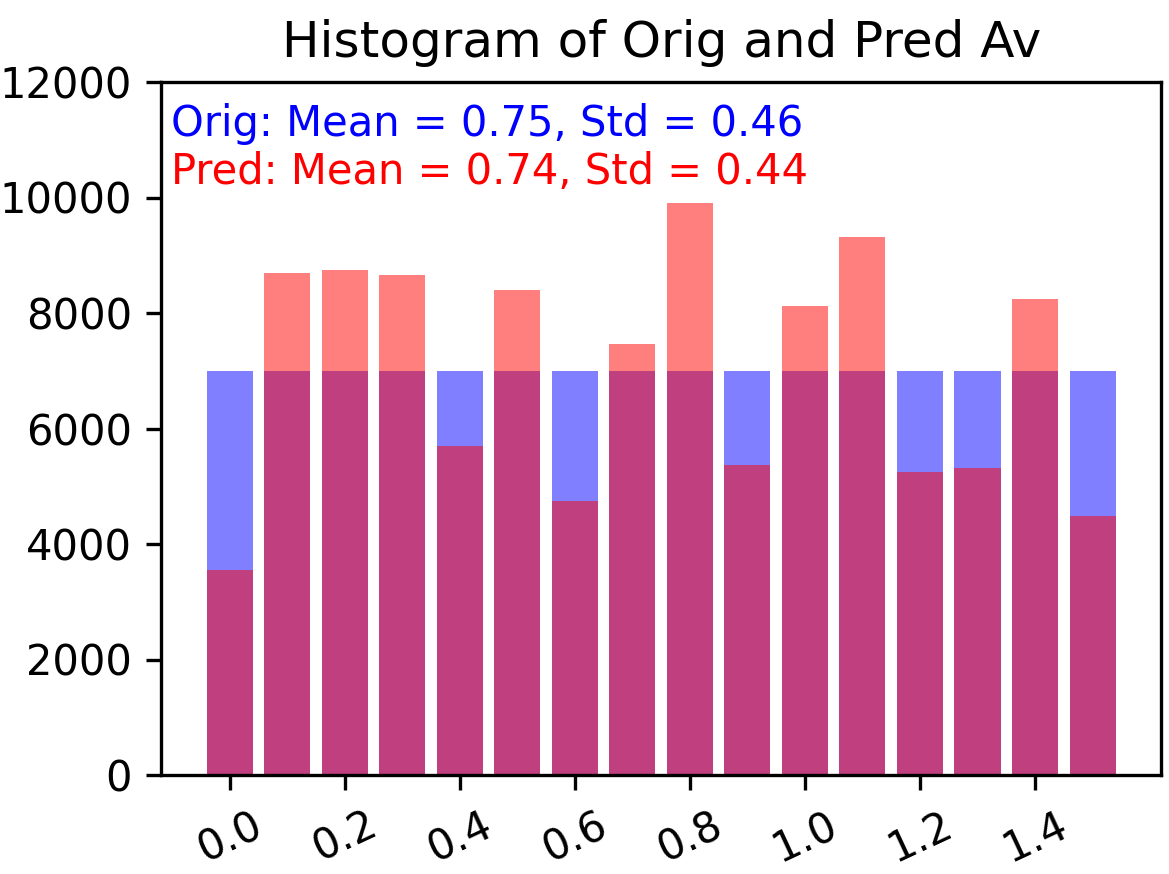}
        \begin{picture}(0,0)
        \put(10,4){\makebox(0,0){\small $A_v$}} 
        \end{picture}
        \begin{picture}(0,0)
        \put(-130,90){\rotatebox{90}{\small Count}} 
        \end{picture}
        \caption{Dust attenuation ($A_v$) distribution histogram.}
        \label{fig:Av_histogram}
    \end{subfigure}
    \hfill
    \begin{subfigure}[t]{0.48\linewidth}  
        \centering
        \includegraphics[width=\linewidth]{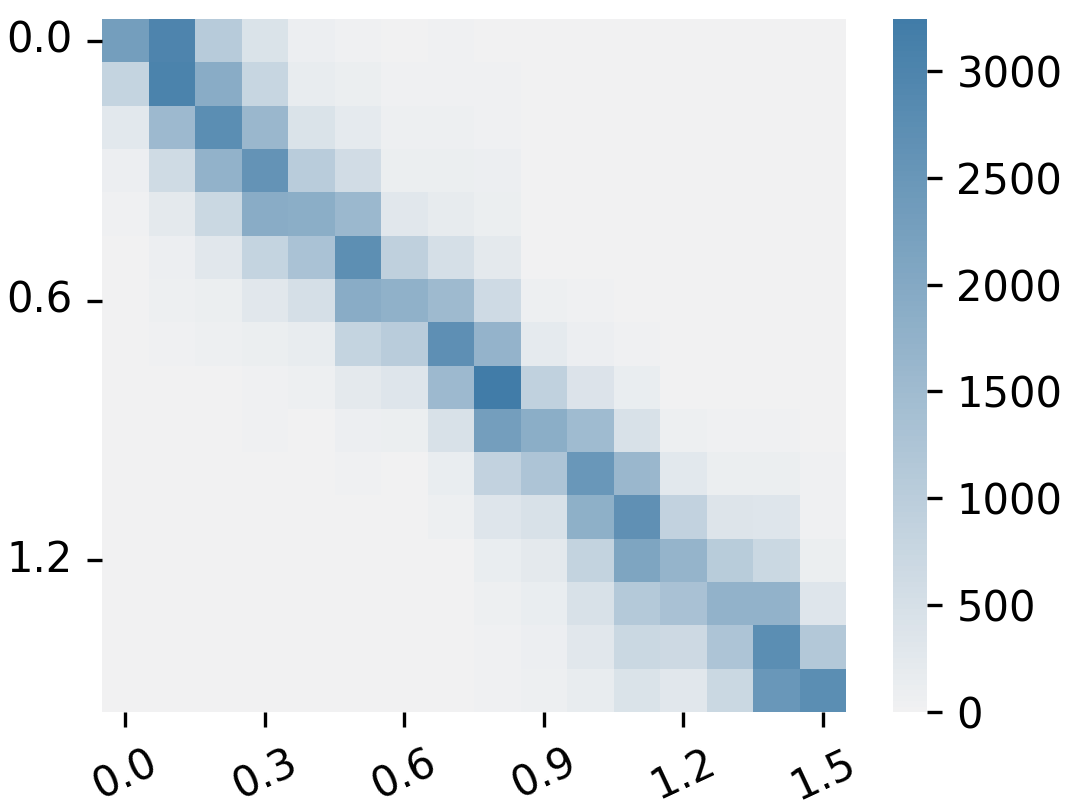}
        \begin{picture}(0,0)
        \put(0,2){\makebox(0,0){\small $A_v$ Pred}} 
        \end{picture}
        \begin{picture}(0,0)
        \put(-130,90){\rotatebox{90}{\small $A_v$ Orig}} 
        \end{picture}
        \caption{Confusion matrix for predicted vs. original dust attenuation.}
        \label{fig:cfm_Av}
    \end{subfigure}
    \caption{Comparison of the original and predicted dust attenuation using a histogram (left) and a confusion matrix (right).}
    \label{fig:Test_Av_Error_stats}
\end{figure}
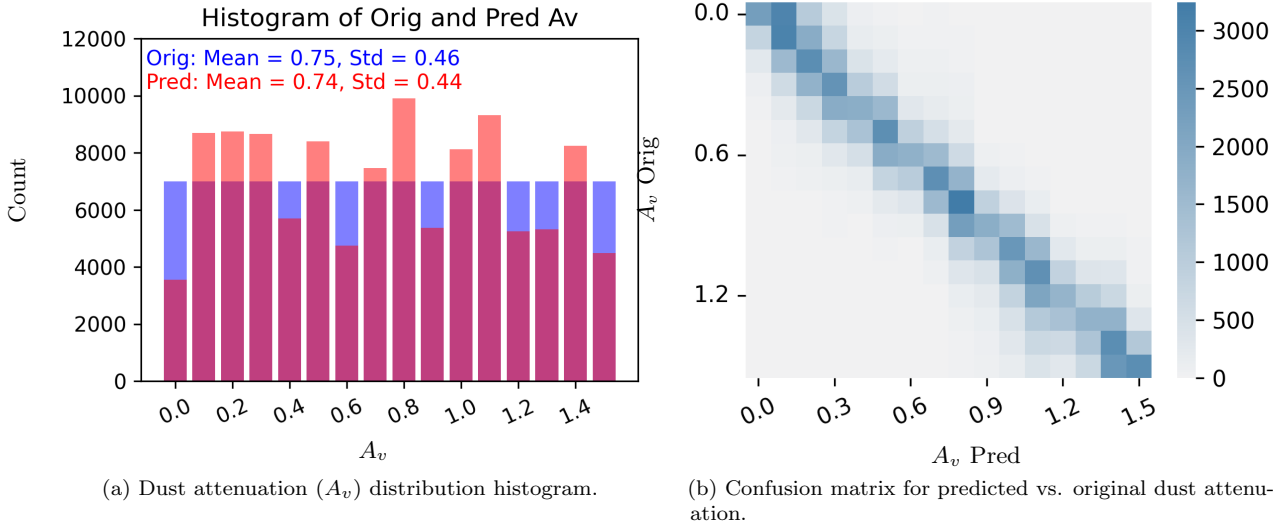

The histogram in Figure \ref{fig:Test_Z_Error_stats} displays the distribution of original and predicted mean metallicities. The close match between the distributions indicates the model's proficiency in identifying metallicity levels in the synthetic data. The bias towards higher metallicities introduced in the training and testing data (Section \ref{Data Pre-processing}) is also prominent, and is more obvious in metallicities predicted by \ShortName. The error between mean metallicity of original dataset and the predictions is only 0.04. The confusion matrix for mean metallicity predictions shows misclassifications for the lowest (-2.27) and highest (0.4) metallicity bins. For other metallicities, a concentration along the diagonal highlights the model's accuracy in classifying the bulk of the metallicity values. 

\subsubsection{\textbf{Dust Attenuation}}

Figure \ref{fig:Test_Av_Error_stats} presents a histogram comparing the distribution of original and predicted dust attenuation values. The original labels show a flat distribution for dust attenuation as the dataset contains an equal number of examples for each attenuation value, unlike age and metallicity where the distribution is Gaussian because of the random combination of templates. The error between the mean extinction of the original dataset and the predictions is 0.01 mag, indicating a negligible bias. However, the performance of \ShortName~in estimating the attenuation is somewhat less optimal when compared to ages and metallicities. The histogram reveals a tendency for the model to favour certain $A_v$ predictions over others. Despite this, the confusion matrix indicates that when the model misclassifies the dust attenuation values, the errors are generally small, with predictions remaining close to the actual values.

\begin{figure}[h!]
\centering
    \includegraphics[width=0.70\linewidth]{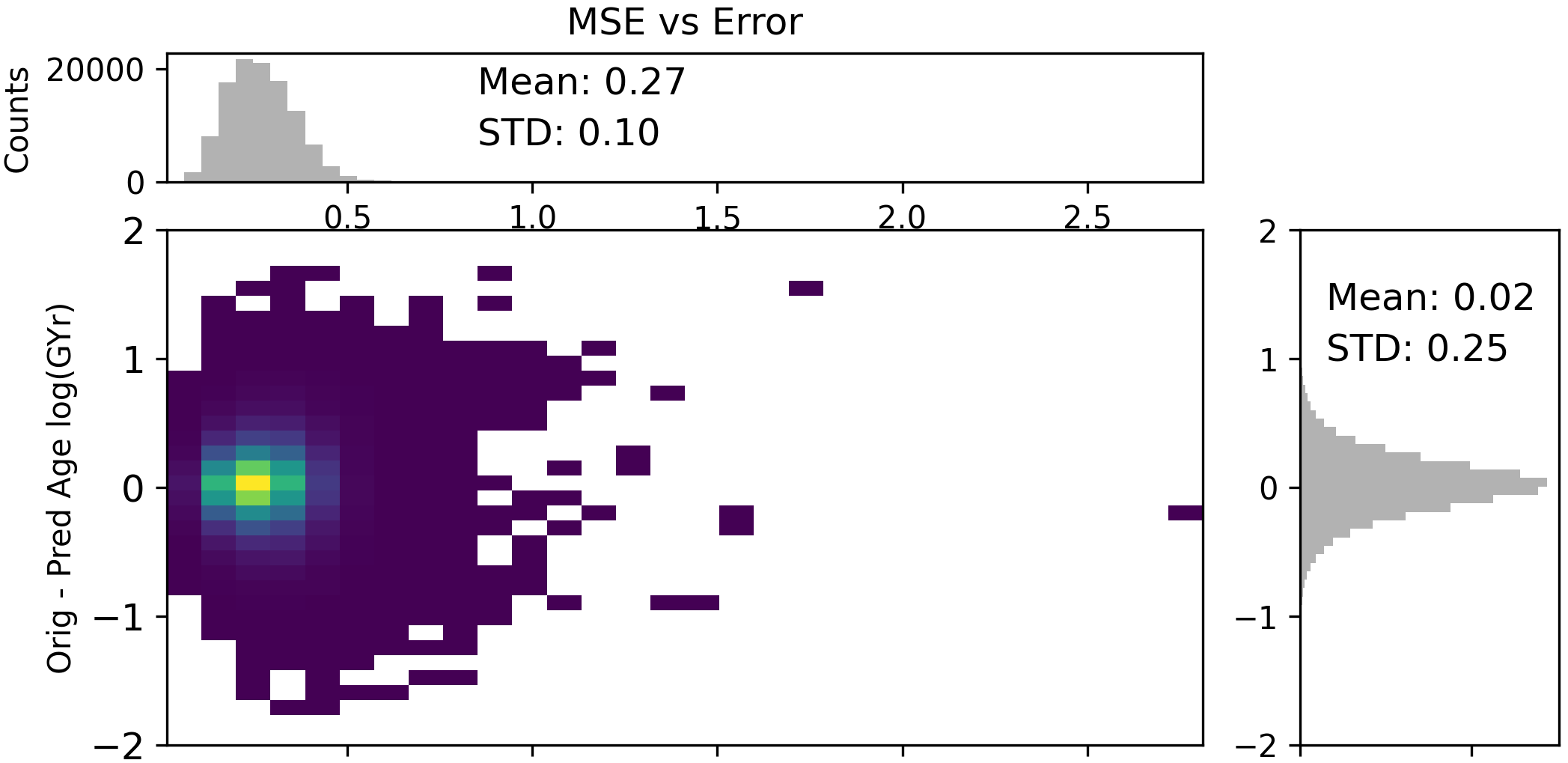}
    \includegraphics[width=0.70\linewidth]{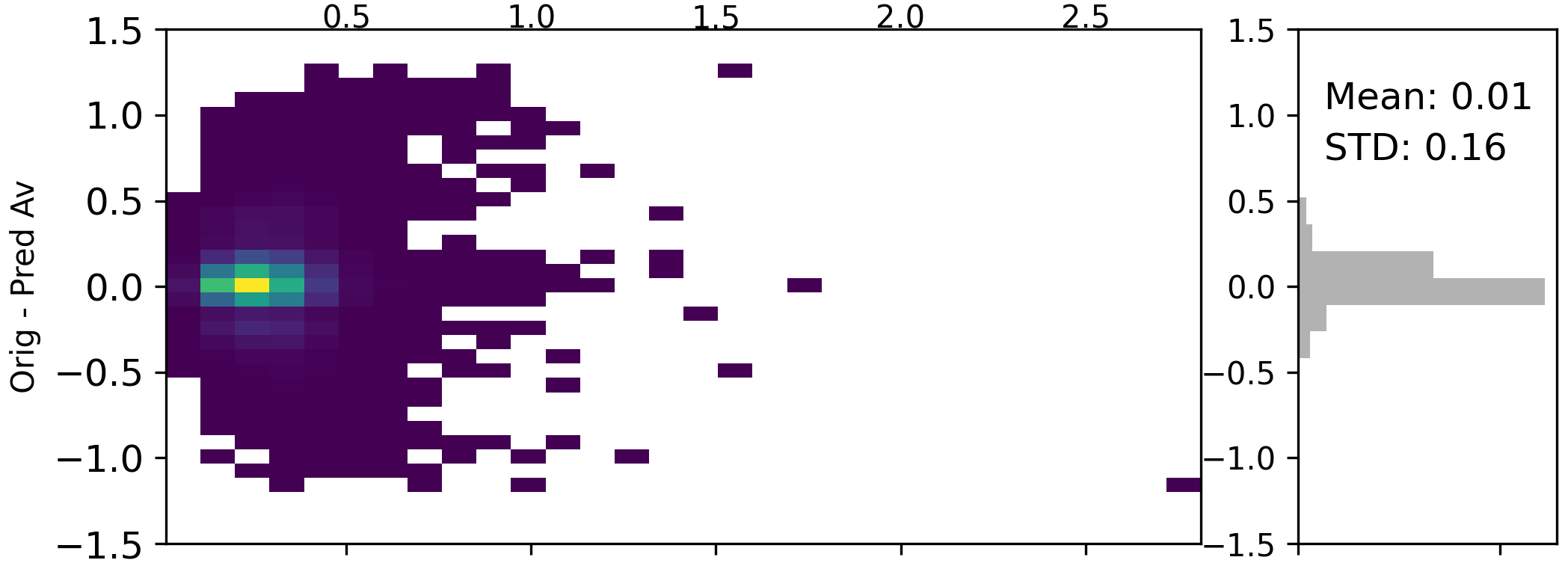}
    \includegraphics[width=0.70\linewidth]{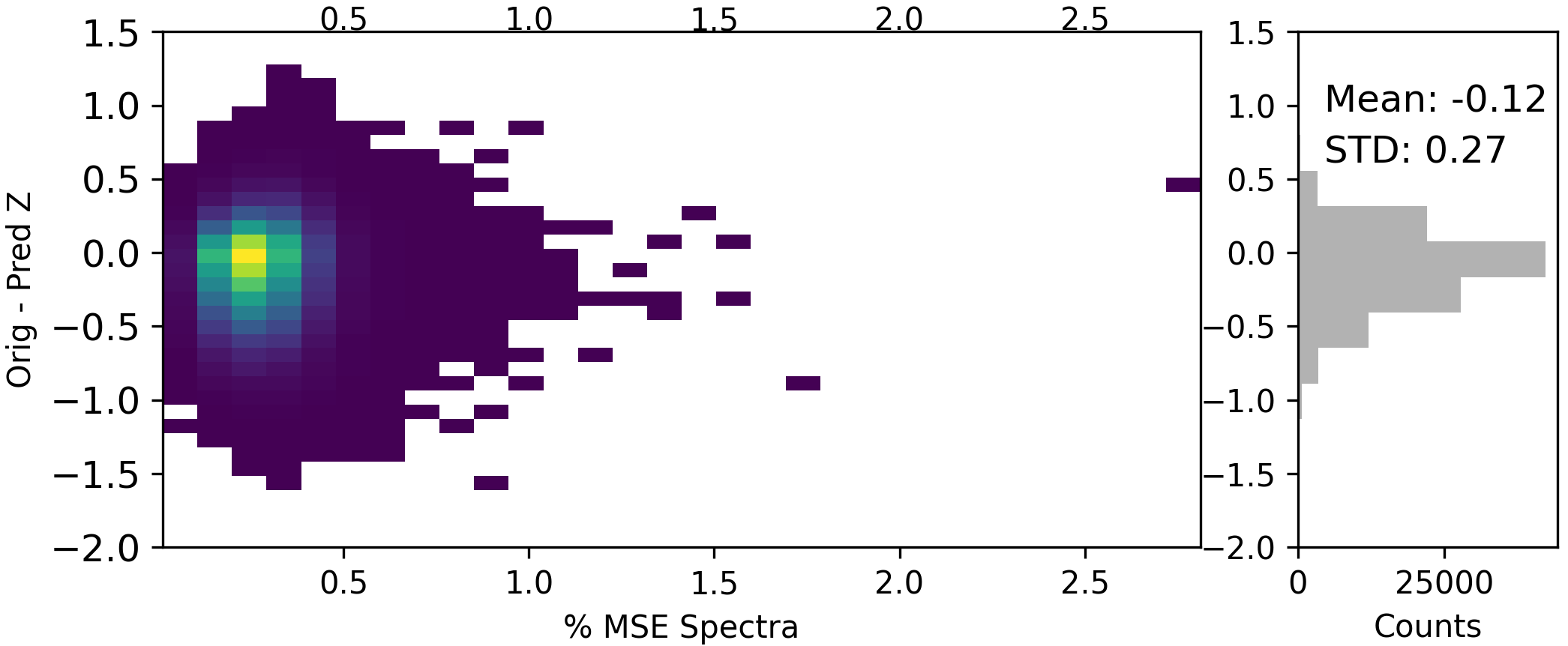}
    
\caption{Errors in age (dex), dust attenuation ($A_v$) and metallicity ([M/H]) predictions plotted against the \% MSE of the test set.}
\label{fig:MSEvsErrors}
\end{figure}

\subsection{\textbf{Spectral Errors}}
\label{p2p}

\subsubsection{\textbf{Mean Squared Error on the Test set}}

For each predicted grid in the test set, we generate the predicted spectrum integrating the E-MILES templates using the corresponding predicted weights within the grid. We then calculate the Percentage Mean Squared Error (MSE) between the predicted spectrum and the input spectrum. The input spectra have been subjected to varying noise levels, while the predicted spectra are derived from the pure E-MILES templates without any added noise. Figure \ref{fig:MSEvsErrors} illustrates the distribution of MSE across the entire test set, which ranges from 0\% to 3\%, with a mean of 0.27\% and standard deviation of 0.10. This figure also shows the relationship between MSE and the resulting errors (original - predicted) in age, metallicity and dust attenuation. The error in age has a mean of 0.02 dex with a standard deviation of 0.25 and is not shown as a function of MSE error. The error in $A_v$ has a mean of 0.01, with a standard deviation of 0.16 and also does not correlate with the MSE error. However, the error in Z indicates that \ShortName~tends to over-predict metallicities as MSE increases. The histogram of Z error distribution reflects this as a mean error (bias) of -0.12 with a standard deviation of 0.27.

\subsubsection{\textbf{Point-to-point Spectral Errors}}
\label{p2p}

In order to assess the performance of \ShortName~in predicting spectra, we evaluated its predictions against both synthetic and PHANGS-MUSE data (NGC4254 and NGC5068). The next section presents a comparison between galactic maps generated by this model and those produced by pPXF. Before delving into this comparison, it is crucial to establish confidence in the quality of the spectral fits generated by the model for the test and the galactic spectra. By comparing the predicted spectra with the actual input spectra, we measured point-to-point discrepancies/residuals and assessed the model's precision. These residuals provide valuable insights into the model's ability to accurately capture the nuances of spectral features across the test dataset, as well as complex observational data.

\begin{figure}[h!]
    \centering
    \includegraphics[width=0.60\linewidth]{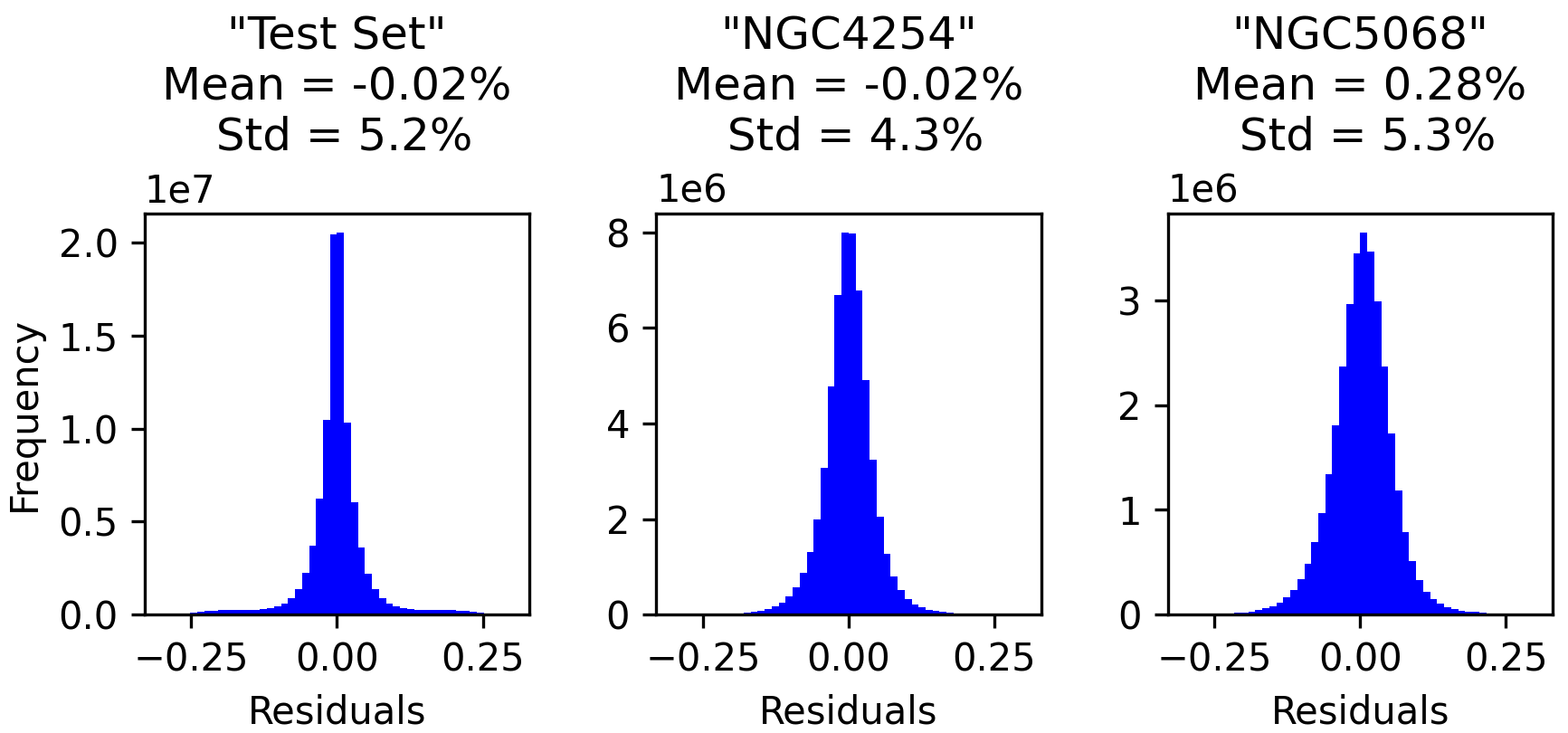}
    \caption{Histograms of the residuals between the input and the predicted spectra.}
    \label{fig:p2p_error_analysis}
\end{figure}

The point-to-point errors were calculated with both input and predicted spectra normalised to a median flux value of 1. For the synthetic test set, which consists of 111,936 templates with 876 samples per template (totalling over 98 million points), the mean point-to-point error between input spectra and fitted spectra is -0.02\%, with a standard deviation of 5.2\%, as shown in Figure \ref{fig:p2p_error_analysis}. This indicates that \ShortName's predictions are, on average, extremely close to the actual values, with minimal deviations. The error distribution in the test set is centred around zero, suggesting that the model effectively captures the underlying features of the spectra used during training. The low standard deviation further supports the model's robustness, as most predictions deviate only slightly from the actual spectra.

In the case of NGC4254, the mean point-to-point error is also -0.02\%, with a standard deviation of 4.3\%. The consistently low mean error indicates a good overall fit for this galaxy. The reduced standard deviation compared to the test set suggests that \ShortName~performs consistently well on this real galactic data. The histogram of errors for NGC4254 shows that the majority of errors are clustered around 0, demonstrating the model's capability to generalise well from synthetic data to real observational data.

For NGC5068, the mean point-to-point error is 0.28\%, with a standard deviation of 5.3\%. The mean error is slightly higher than that of the NGC4254, because of relatively poor SNR and the velocity gradient. The standard deviation is comparable to that of test set, suggesting consistent performance.

Overall, \ShortName~demonstrates high accuracy and robustness, with mean residual errors of less than 0.3\% and standard deviations of around 5\% across both synthetic and galactic data. The slightly higher errors for NGC5068 compared to the test set can be attributed to the inherent complexity and noise present in observational data. Nevertheless, the model's ability to maintain a low mean error and standard deviation in real-world applications underscores its potential for effective galactic characterisation. These results validate the model's capacity to generalise from synthetic training data to real observational data, paving the way for more efficient and accurate astronomical studies.

\subsection{\textbf{Comparison with pPXF maps from PHANGS-MUSE}}

This section provides the comparison between the maps generated by \ShortName~ and those produced by pPXF as part of the PHANGS pipeline \citep[see][]{2022A&A...659A.191E}, using MUSE data cubes for two galaxies from the PHANGS-MUSE survey. For this comparison, we focus exclusively on light-weighted (LW) spectra for both galaxies, in the range of 4750–7100~\AA. The spectra were pre-processed to correct for redshift (assuming a single redshift per galaxy) and to filter out bright lines using median filters. Additionally, major emission lines were masked as described in Section \ref{Masking}.

There are several key distinctions between the two methods being compared. pPXF in the PHANGS-MUSE pipeline is run twice: the first iteration determines extinction, which is then fixed in the second iteration for extracting stellar properties. A multiplicative polynomial is applied in this step to correct residual systematics in flux calibration, and does not employ regularization. In contrast, our approach does not include such a polynomial correction, and uses a single iteration for extracting all three stellar properties. Additionally, the age-metallicity grid used by pPXF (6 × 13) is of lower resolution than our model’s grid (12 × 53) (see Section \ref{Dataset Preparation}). For further details on pPXF’s configuration in PHANGS-MUSE, see \citep{2022A&A...659A.191E}.

\clearpage
\subsubsection{\textbf{Ages}}

\begin{figure}[b!]
    \centering
    \begin{subfigure}[b]{0.80\linewidth}
        \centering
        \includegraphics[width=0.80\linewidth]{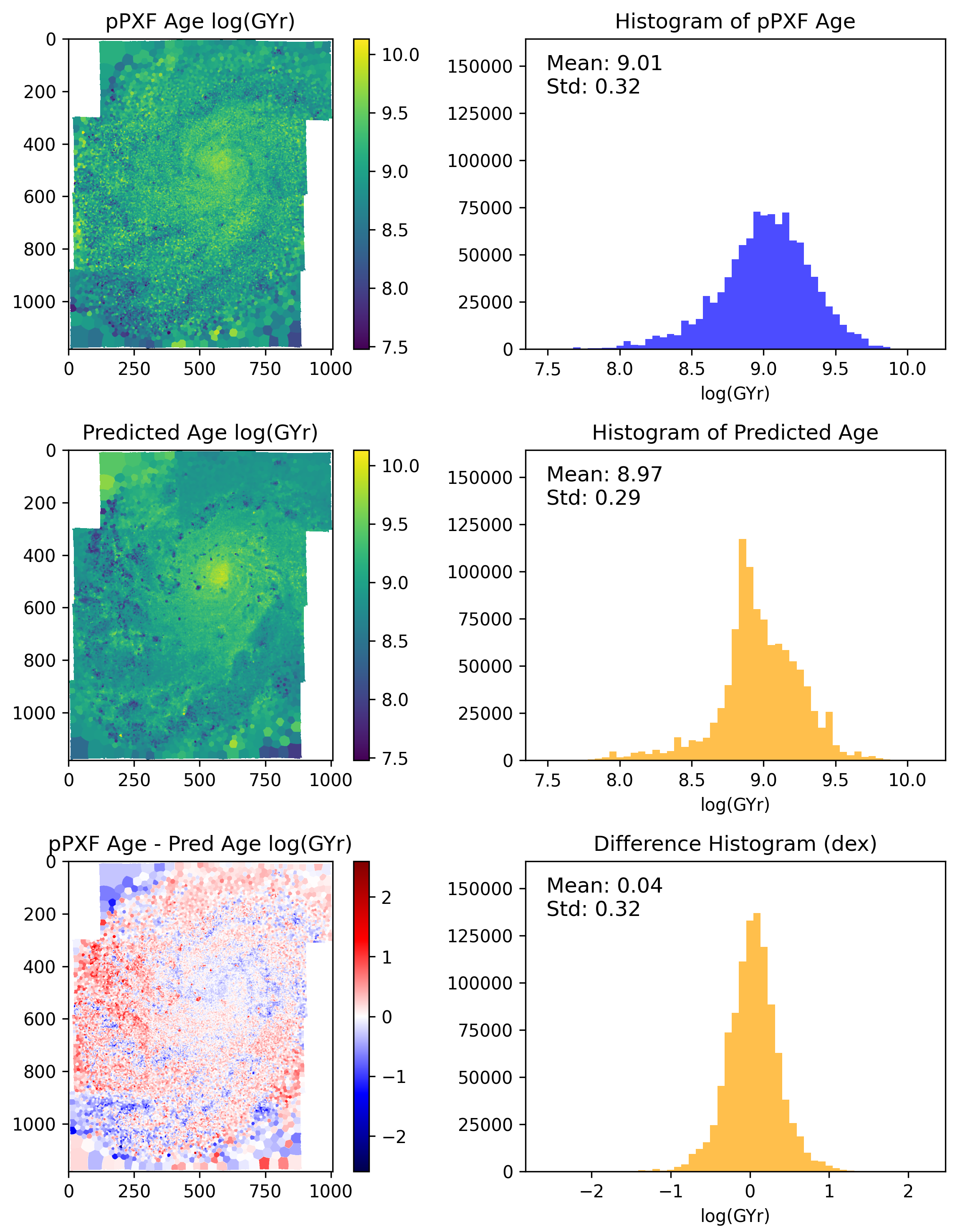}
        \caption{pPXF map (top), predicted map (middle), and difference map (bottom).}
        \label{fig:4254_Age_Comparison}
        
    \end{subfigure}
    \begin{subfigure}[b]{0.80\linewidth}
        \centering
        \includegraphics[width=0.80\linewidth]{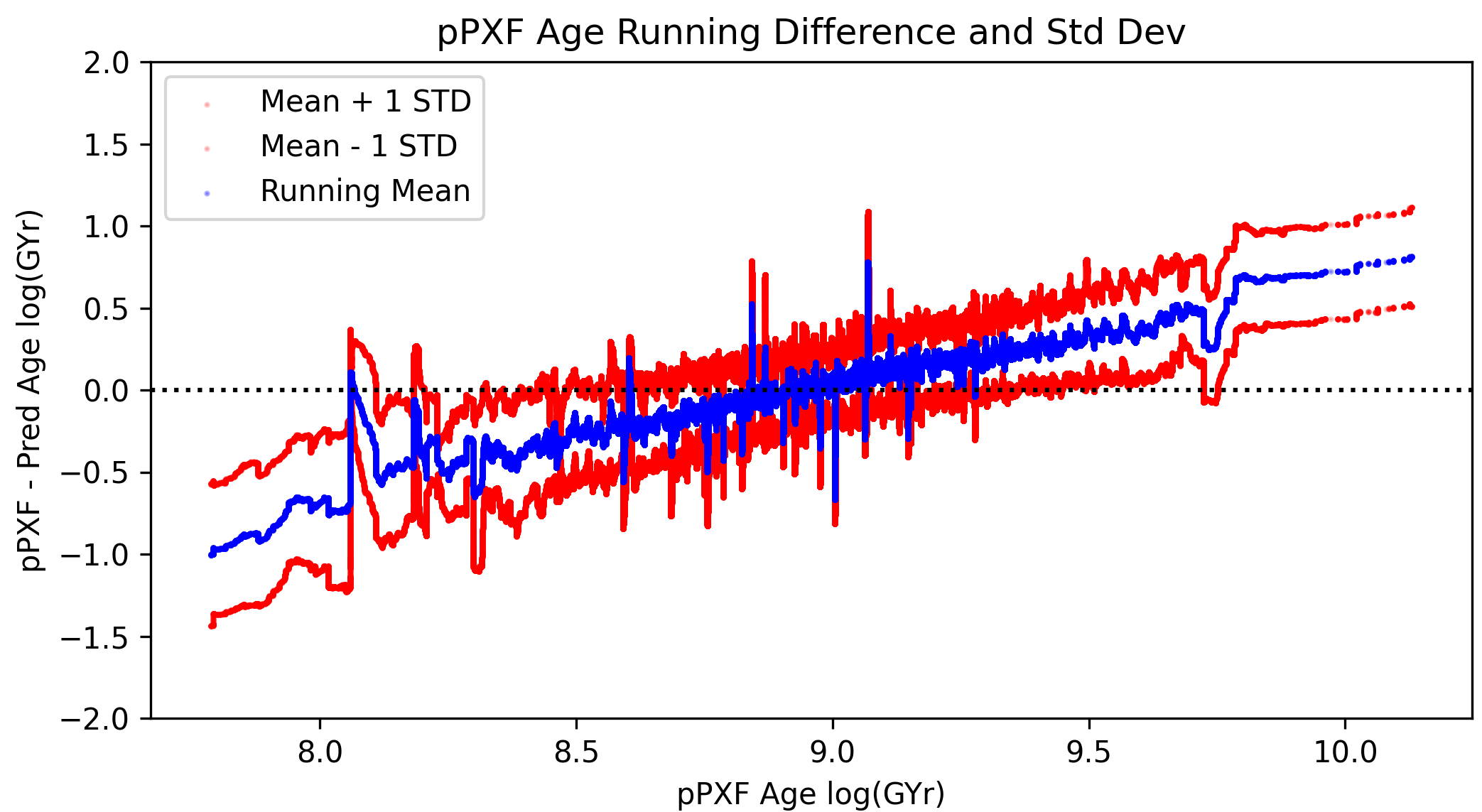}
        \caption{The running difference against pPXF age (dex).}
        \label{fig:4254_Age_Error}
    \end{subfigure}
    \caption{Age maps and running difference for NGC4254.}
\end{figure}

\begin{figure}[b!]
    \centering
    \begin{subfigure}[b]{0.80\linewidth}
        \centering
        \includegraphics[width=0.80\linewidth]{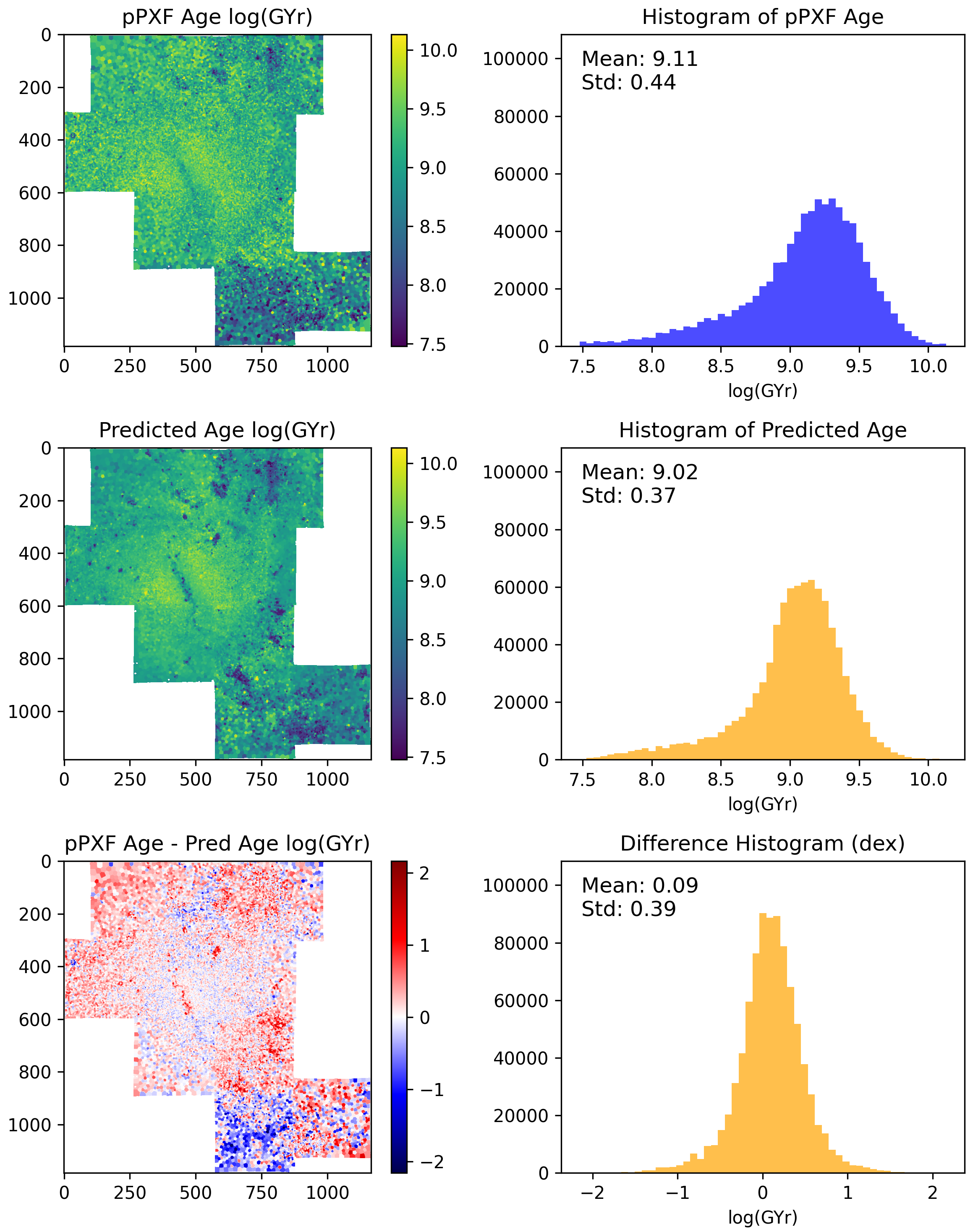}
        \caption{pPXF map (top), predicted map (middle), and difference map (bottom).}
        \label{fig:5068_Age_Comparison}
    \end{subfigure}
    \begin{subfigure}[b]{0.80\linewidth}
        \centering
        \includegraphics[width=0.80\linewidth]{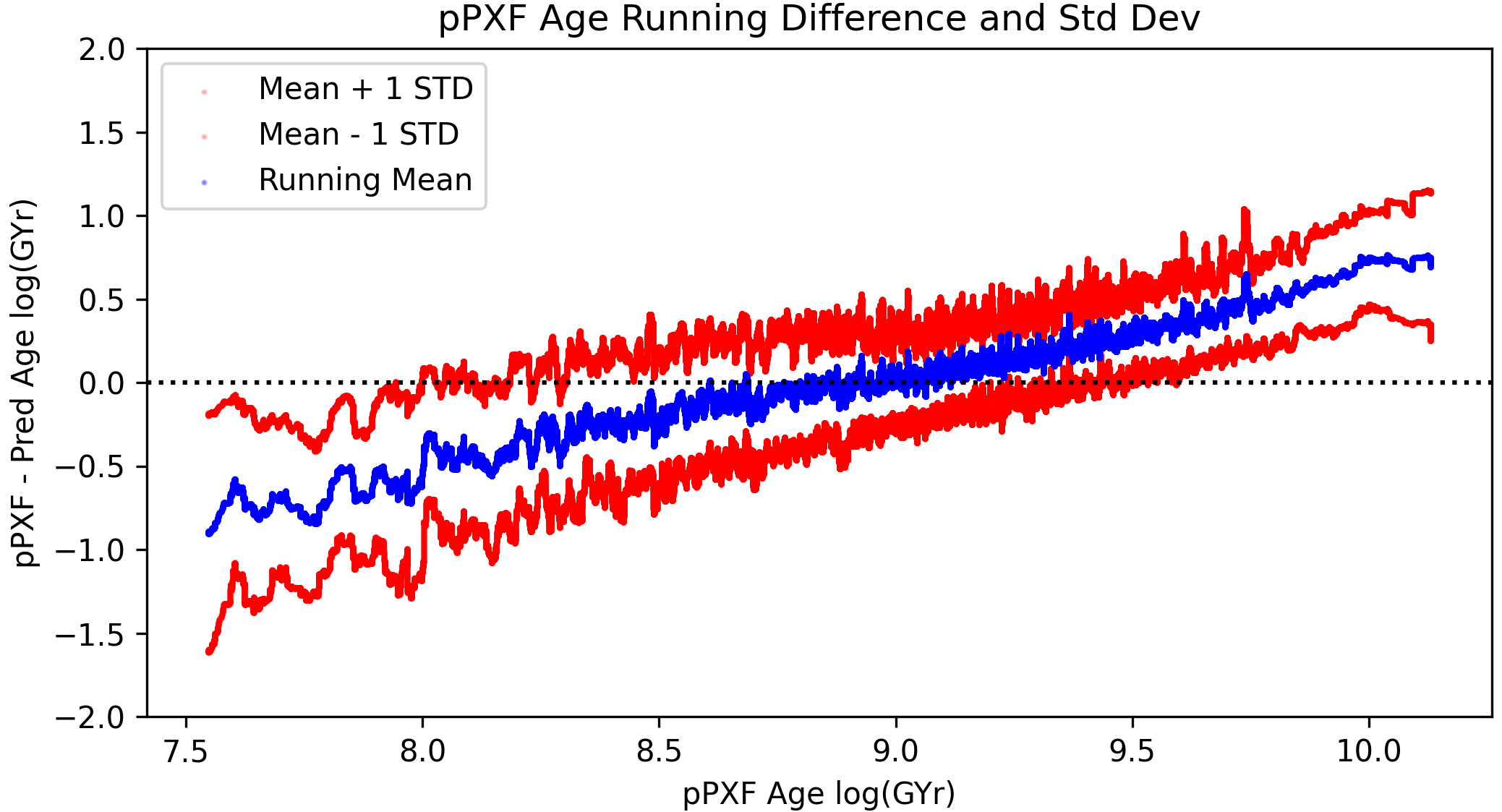}
        \caption{The running difference against pPXF age (dex).}
        \label{fig:5068_Age_Error}
    \end{subfigure}
    \caption{Age maps and running difference for NGC5068.}
\end{figure}

From the age maps of both NGC4254 and NGC5068 (Figures \ref{fig:4254_Age_Comparison} and \ref{fig:5068_Age_Comparison}), it is clear that the proposed approach results in a smoother, less noisy, and higher local dynamic range of stellar population distribution compared to PHANGS-MUSE. The probability distribution function (PDF) for the \ShortName~is narrower, indicating tighter constraints on the predicted ages, and some systematic bias leaning away from lowest and highest ages in the training data. Despite the differences in structure, the mean ages between the two methods are consistent, with mean differences of 0.04 and 0.09 dex. The standard deviations of predicted age maps are 0.29 dex for NGC4254 and 0.37 dex for NGC5068, both lower than their respective pPXF maps. 

The difference maps between the two methods exhibit noticeable structure, suggesting that at least part of the discrepancy arises from how pPXF and \ShortName~handle different spectral features. This trend becomes more apparent when examining the running difference between the two age maps (Figures \ref{fig:4254_Age_Error} and \ref{fig:5068_Age_Error}), where the largest deviations occur for ages that significantly diverge from the average population age. For both galaxies, \ShortName~tends to underestimate older ages and overestimate younger ages, biasing the predictions toward the mean age in the training data.

\clearpage

\subsubsection{\textbf{Metallicities}}

\begin{figure}[b!]
    \centering
    \begin{subfigure}[b]{0.80\linewidth}
        \centering
        \includegraphics[width=0.80\linewidth]{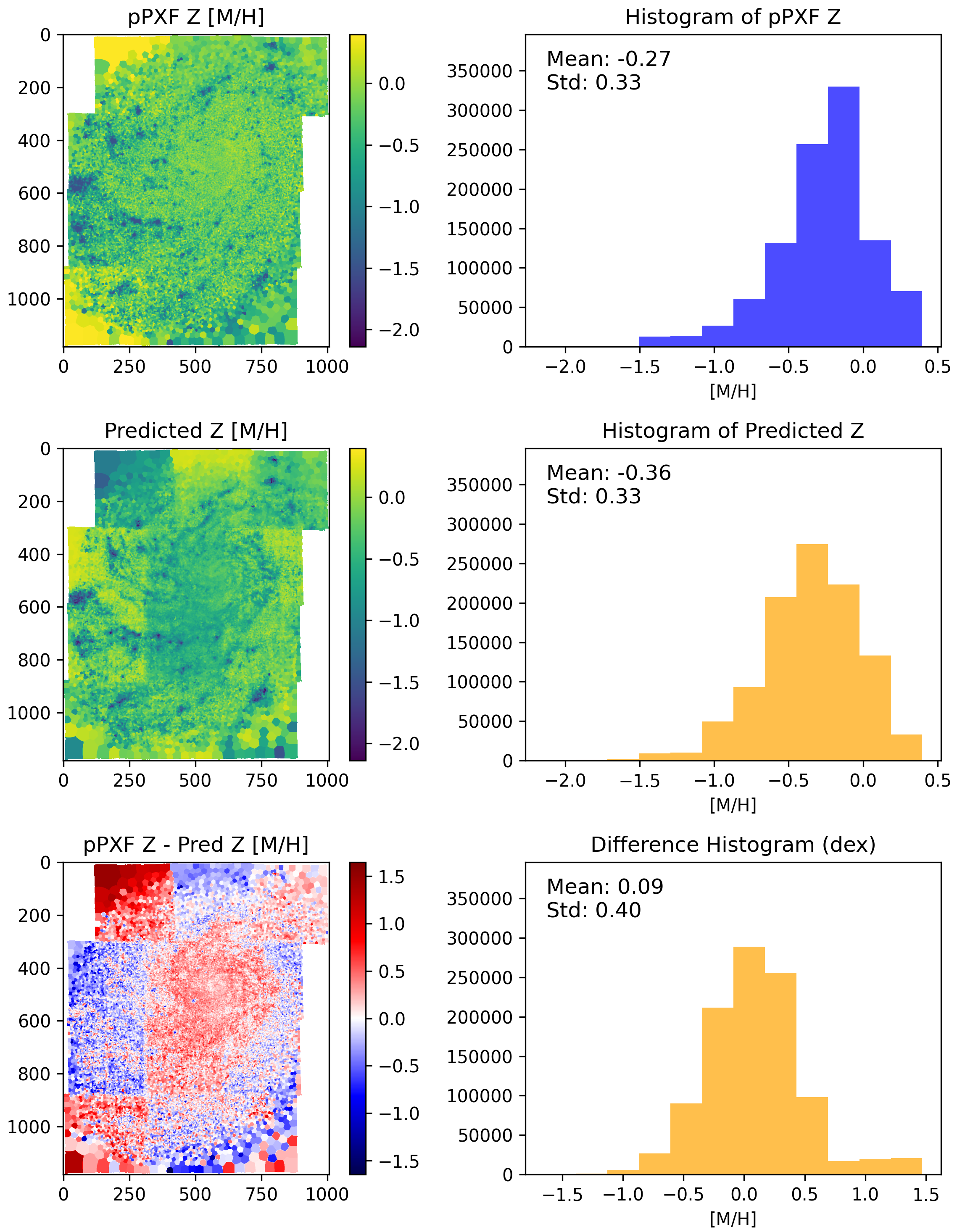}
        \caption{pPXF map (top), predicted map (middle), and difference map (bottom). }
        \label{fig:4254_Z_Comparison}
    \end{subfigure}
    \begin{subfigure}[b]{0.80\linewidth}
        \centering
        \includegraphics[width=0.80\textwidth]{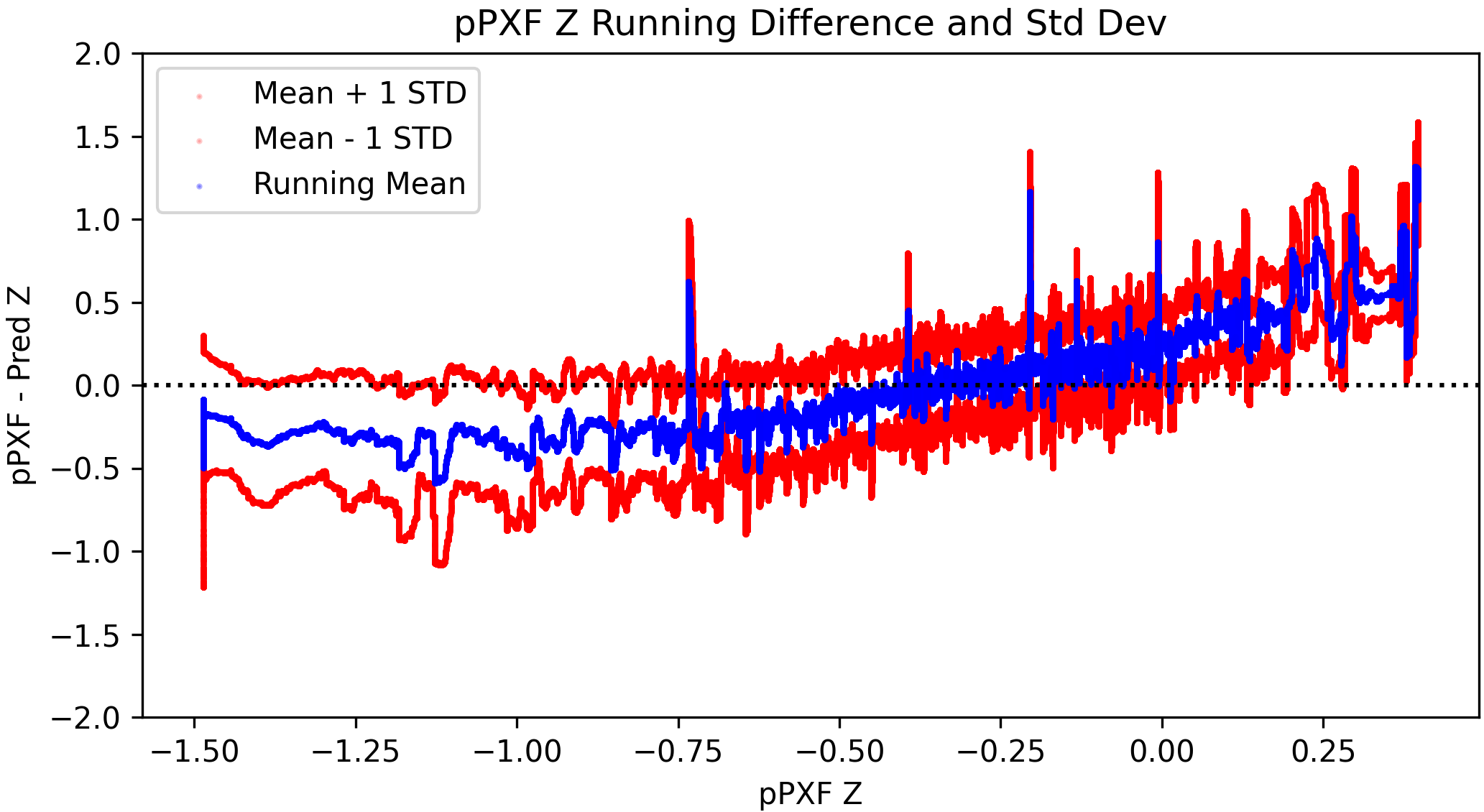}
        \caption{The running difference against pPXF metallicity [M/H] values.}
        \label{fig:4254_Z_Error}
    \end{subfigure}
    \caption{Metallicity maps and running difference for NGC4254.}
\end{figure}

\begin{figure}[b!]
    \centering
    \begin{subfigure}[b]{0.80\linewidth}
        \centering
        \includegraphics[width=0.80\linewidth]{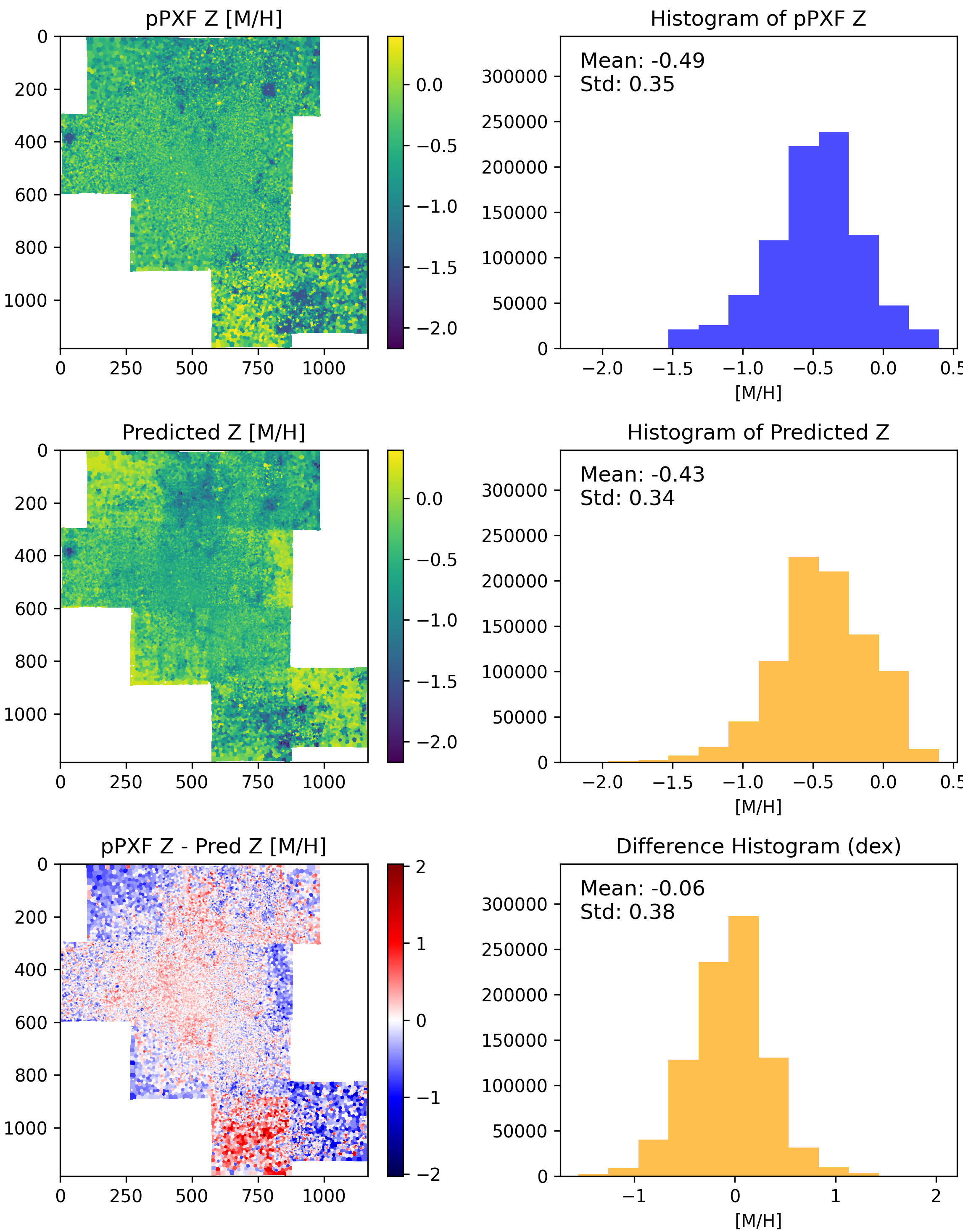}
        \caption{pPXF map (top), predicted map (middle), and difference map (bottom).}
        \label{fig:5068_Z_Comparison}
    \end{subfigure}
    \begin{subfigure}[b]{0.80\linewidth}
        \centering
        \includegraphics[width=0.80\linewidth]{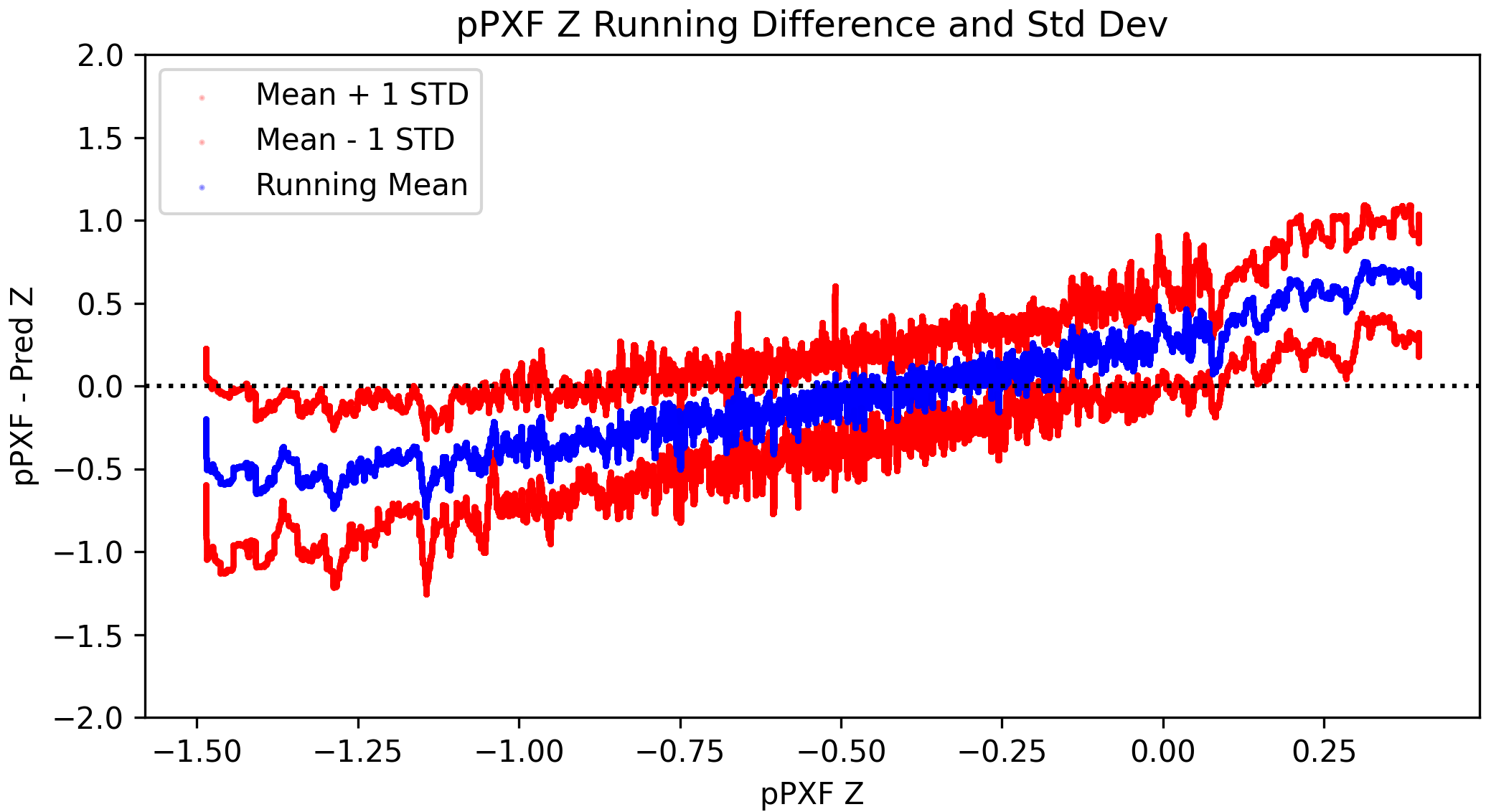}
        \caption{The running difference against pPXF metallicity [M/H] values.}
        \label{fig:5068_Z_Error}
    \end{subfigure}
    \caption{Metallicity maps and running difference for NGC5068.}
\end{figure}

For both galaxies, the metallicity maps generated by \ShortName~display more pronounced structure compared to pPXF (Figures \ref{fig:4254_Z_Comparison} and \ref{fig:5068_Z_Comparison}). This suggests that the proposed model is more affected by the spatial variations within the galaxies compared to PHANGS-MUSE. While the overall mean metallicity are consistent within 0.09 dex for NGC4254 and 0.06 dex for NGC5068, there are distinct differences in how the methods allocate metallicity across the maps.

As already seen in the case of stellar ages, when it comes to metallicities \ShortName~ overestimates stellar metallicities with respect to pPXF for values below the average value observed in the target galaxy, and underestimates its value for higher metallicities (see Figures \ref{fig:4254_Z_Error} and \ref{fig:5068_Z_Error}) 

Interestingly, it appears that \ShortName~ assign the normalisation offsets in the data due to issues with sky subtractions to metallicity, contrary to what was observed in the case of pPXF, where normalisation issues between MUSE pointings mainly affect the extinction estimates. 
This distribution of residuals, combined with the structural differences in the maps, highlights how the proposed approach responds differently to the spectral features of the galaxies.

\clearpage

\subsubsection{\textbf{Dust Attenuation}}

\begin{figure}[b!]
    \centering
    \begin{subfigure}[b]{0.80\linewidth}
        \centering
        \includegraphics[width=0.80\linewidth]{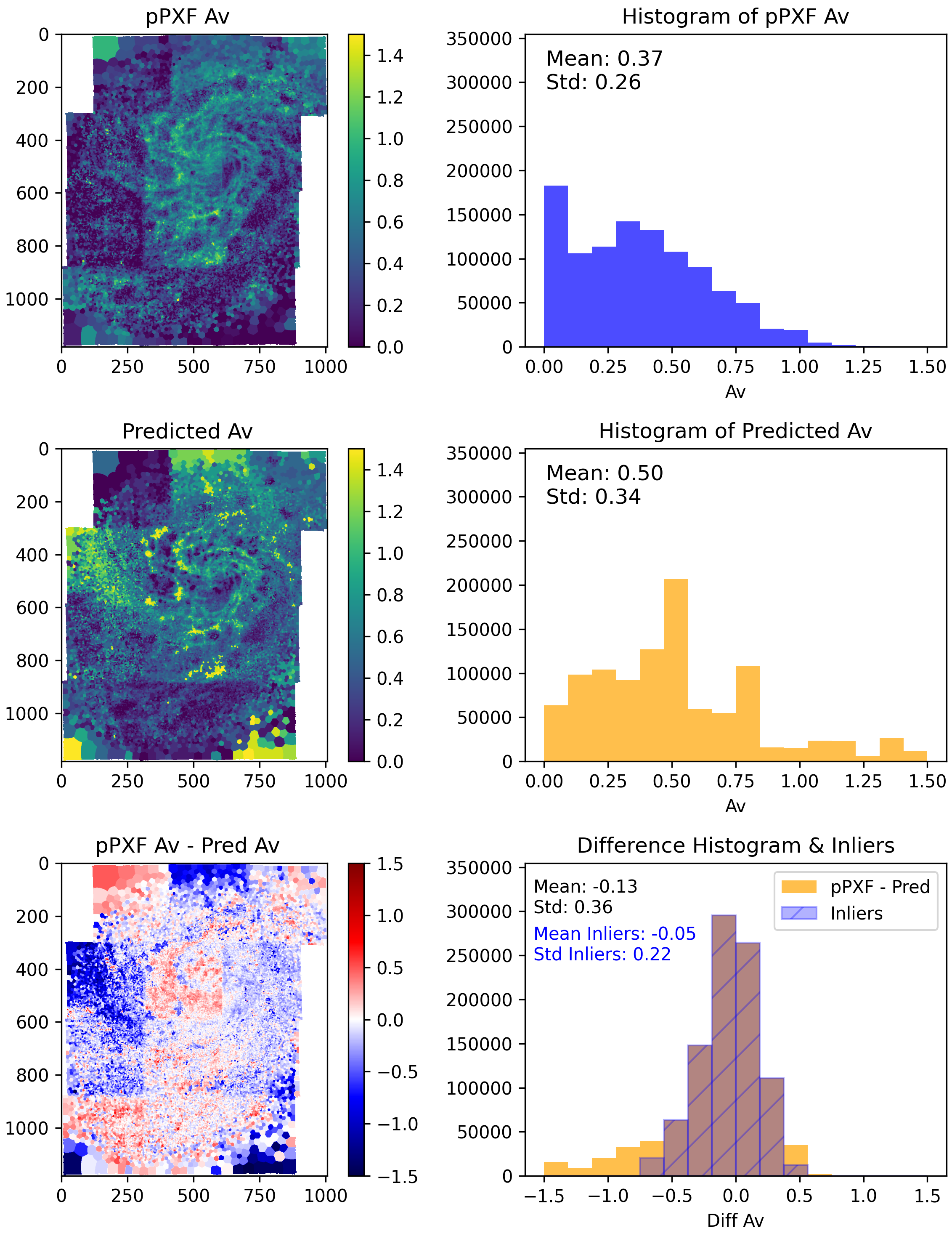}
        \caption{pPXF map (top), predicted map (middle), and difference map (bottom). }
        \label{fig:4254_Av_Comparison}
    \end{subfigure}
    \begin{subfigure}[b]{0.80\linewidth}
        \centering
        \includegraphics[width=0.80\textwidth]{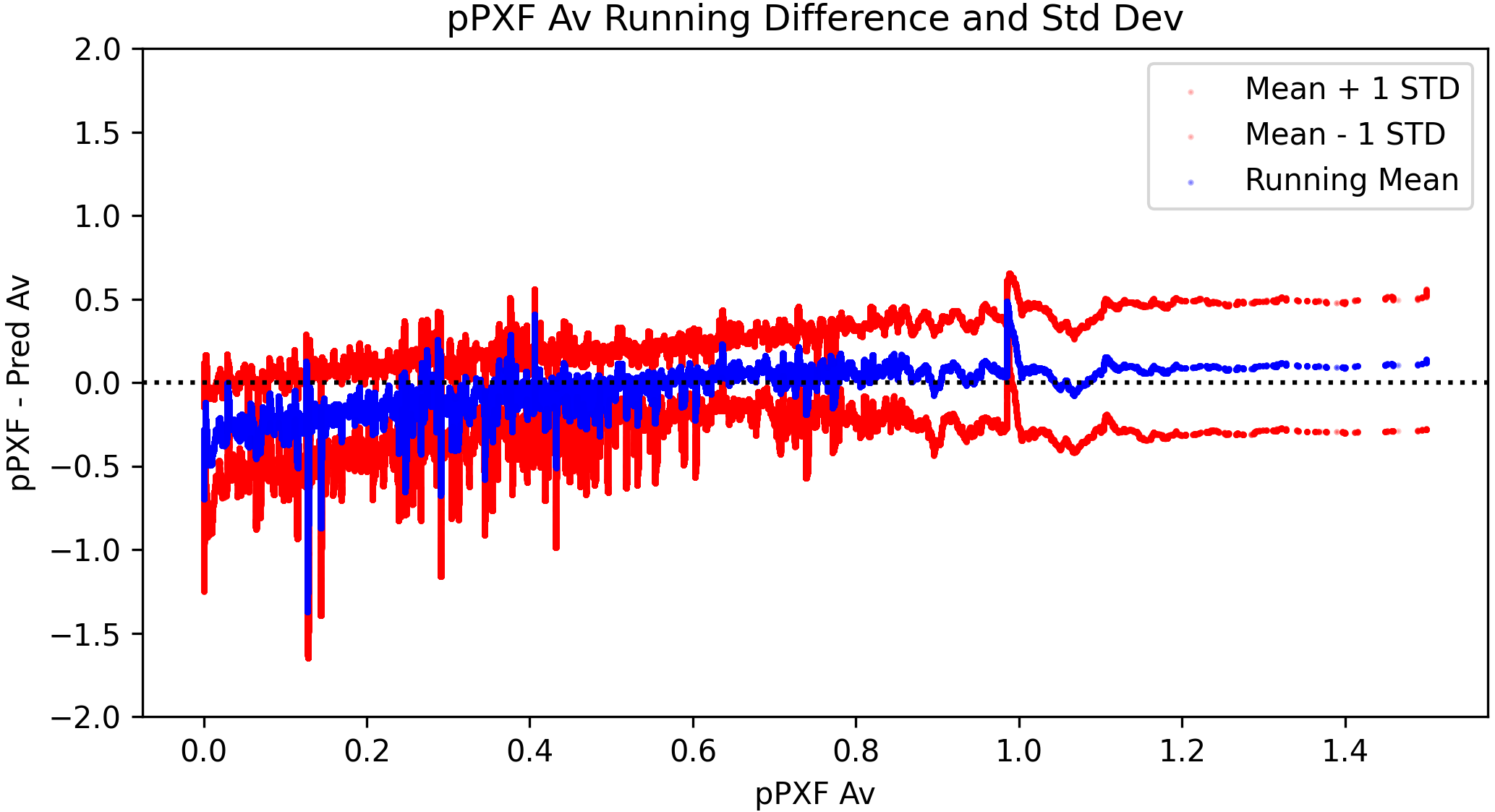}
        
        \caption{The running difference against pPXF dust attenuation values.}
        \label{fig:4254_Av_Error}
    \end{subfigure}
    \caption{Dust attenuation maps and running difference for NGC4254.}
\end{figure}

\begin{figure}[b!]
    \centering
    \begin{subfigure}[b]{0.80\linewidth}
        \centering
        \includegraphics[width=0.80\linewidth]{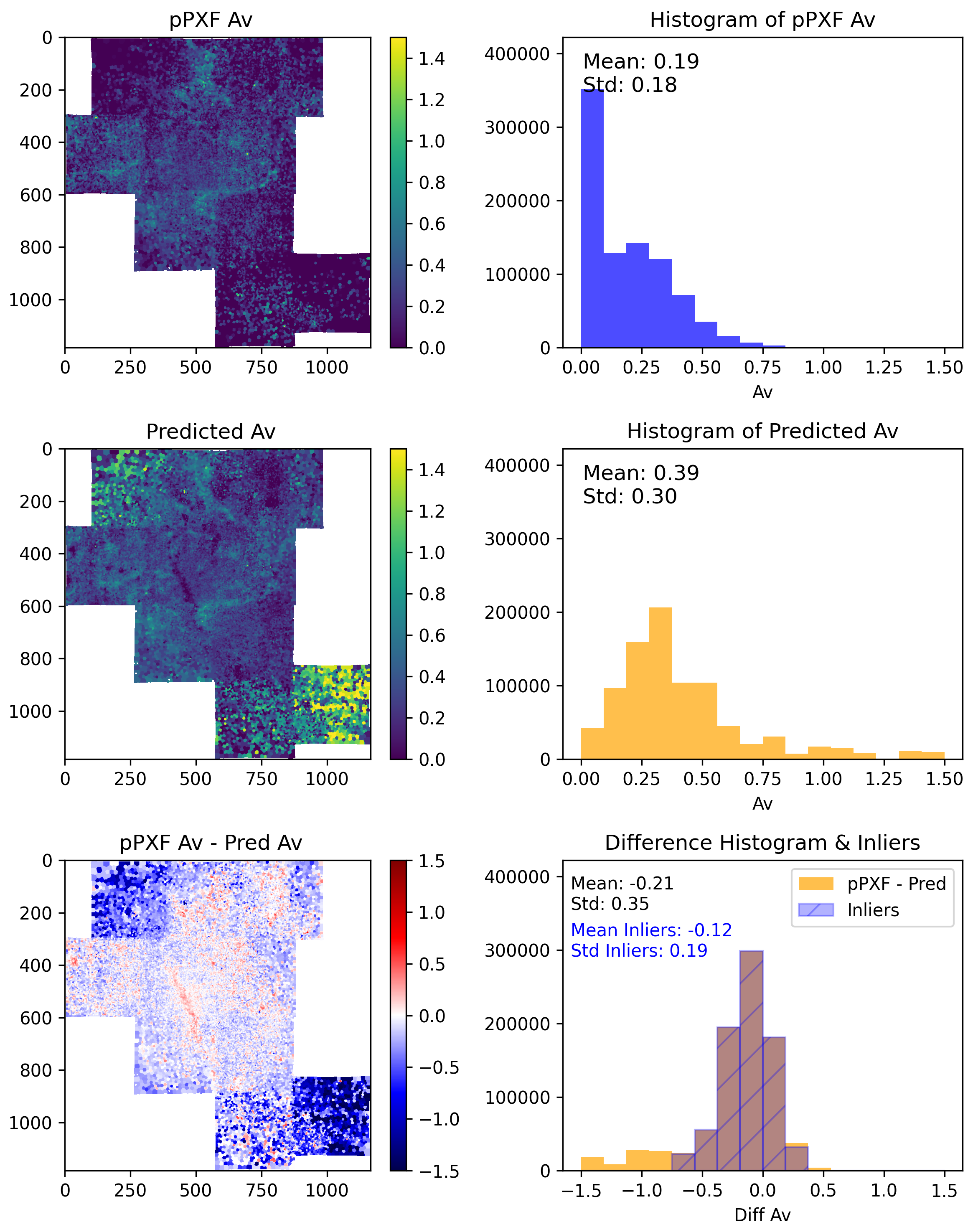}
        \caption{pPXF map (top), predicted map (middle), and difference map (bottom). }
        \label{fig:5068_Av_Comparison}
    \end{subfigure}
    \begin{subfigure}[b]{0.80\linewidth}
        \centering
        \includegraphics[width=0.80\textwidth]{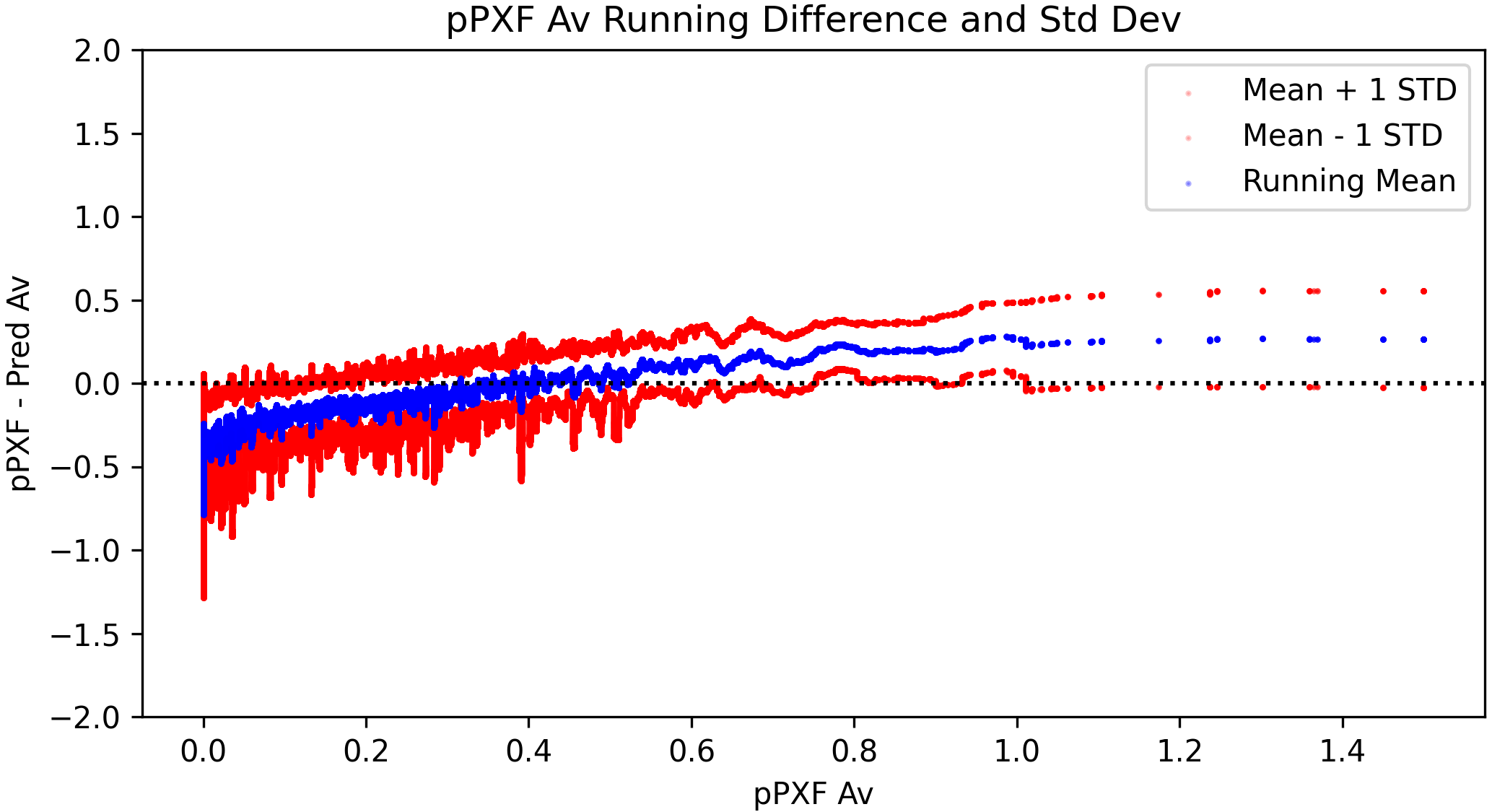}
        \caption{The running difference against pPXF dust attenuation values.}
        \label{fig:5068_Av_Error}
    \end{subfigure}
    \caption{Dust attenuation maps and running difference for NGC5068.}
\end{figure}

The dust attenuation ($A_v$) maps for NGC4254 and NGC5068 exhibit both similarities and key differences (Figures \ref{fig:4254_Av_Comparison} and \ref{fig:5068_Av_Comparison}). In both galaxies, \ShortName~successfully captures the overall dust structure. However, for NGC4254, the difference histogram reveals a mean dust attenuation error of -0.13 mag, with the residuals from the sky background subtraction contributing significantly to this error. Removing the outliers (as shown by the blue shaded area) improves the mean difference to -0.05 mag, with a reduced standard deviation of 0.22 mag. The differences between the two methods in this case highlight how the proposed approach is more sensitive to poor SNR and sky residuals, which particularly affect the dust attenuation predictions.

For NGC5068, the sky residual’s impact on the dust attenuation maps is notably smaller compared to NGC4254. However, the \ShortName~predicted dust attenuation map exhibits a significant offset with a mean error of -0.21 mag. This offset is most prominent in regions of the galaxy with poor SNR. Excluding the outliers from these low-SNR regions leads to substantial improvements in both the mean and standard deviation of the difference profile. Despite the presence of outliers towards the galaxy’s edges, the overall dust structures remain quite similar between the two approaches, reflecting the model’s robustness in dealing with extinction, even in regions where data quality is less reliable.

As in the case of stellar metallicities, some of the systematic differences observed in the maps generated by pPXF and \ShortName~can be attributed to the inherent effects of sky subtraction and filtering in the mosaic data. The residuals from sky subtraction in the reduction process of PHANGS-MUSE \citep{pessa2023resolved}, and the median filtering of bright lines for \ShortName~can introduce subtle artefacts and variations that are processed differently by each method. These discrepancies can lead to distinct differences in the resulting maps. While pPXF applies corrections that may suppress or amplify specific features, \ShortName~may interpret and propagate these residuals differently, leading to the systematic variations observed between the maps.

\section{\textbf{Discussion}}

\subsection{\textbf{Advantages over Traditional Methods}}
This approach offers significant speed advantages over traditional methods. Using \ShortName, the spectral fits for NGC4254 (with 64,985 spectral bins) and NGC5068 (with 37,690 spectral bins) were completed in just 245 and 148 seconds, respectively. In contrast, pPXF takes up to 11 seconds to fit a single spectrum \citep{woo2024comparison}. With \ShortName, more than 250 spectral fits are generated per second, equating to less than 4 milliseconds per fit. This represents a performance boost of over 2,750 times compared to traditional methods. Such a dramatic improvement in efficiency not only accelerates galactic studies but also enables the analysis of much larger datasets that were previously too time-consuming to handle. 

Although GPUs like the NVIDIA RTX 4090 consume significantly more power than CPUs, their extreme speed advantage makes them far more energy-efficient for large-scale spectral fitting. The RTX 4090, with a power draw of 450W, processes 250 spectra per second, resulting in an energy consumption of 1.8W per spectrum. In contrast, pPXF running on a single core of an Intel i9-13900K (which consumes 10.5W per core) processes just 0.09 spectra per second, requiring 117W per spectrum.

When applied to the NGC4254 dataset, this translates to a total energy consumption of 117 kW (or 30.6 Wh) for \ShortName~versus 7.6 MW (or 2,085 Wh) for pPXF—a staggering 68-fold improvement in energy efficiency. Even under conservative estimates, this efficiency gain makes \ShortName~not only a computationally superior choice but also a significantly more sustainable one, reducing both computational cost and environmental impact in large-scale astronomical analyses.

\subsection{\textbf{Limitations and Potential Improvements}}

The model's reliance on synthetic E-MILES data presents a limitation, as synthetic data, even with added noise, cannot fully capture the complexities and nuances of real astronomical observations. Additionally, the model's performance is constrained by the rest-frame wavelength range of 4749.37~\AA~to 7100.49~\AA, limiting its applicability to data outside this range. Another challenge lies in the imbalance within the training dataset, which leads to a Gaussian-like distribution of mean age and metallicity. This imbalance causes \ShortName~to over-predict lower values and under-predict higher values. Future work will address these issues by creating a more balanced training dataset and implementing strategies such as adaptive loss functions or re-weighted sampling to mitigate biases at parameter boundaries.

\ShortName~is optimized for PHANGS-MUSE data, necessitating adjustments for use with other instruments and surveys. With fine-tuning, it can accommodate different spectral resolutions and metallicity distributions across surveys. Additionally, while the current model predicts a 2-D age-metallicity grid with a separate dust attenuation value, extending it to a fully integrated 3-D age-metallicity-extinction grid could improve the interpretation of mixed stellar populations with varying dust attenuation.

The decision to test face-on galaxies was intentional, as their simpler velocity structures minimize the impact of kinematics on spectral fitting. This allowed us to refine the model’s ability to extract stellar population properties efficiently. However, GalProTE does not account for velocity dispersion or kinematic broadening, which could introduce biases in systems with strong velocity gradients. Future improvements may explore incorporating kinematic priors or leveraging ML-based velocity modelling techniques to extend the model's applicability to more dynamically complex galaxies.

Integrating deep learning techniques into astronomical research significantly improves analysis efficiency, enabling more detailed insights into galaxy formation and evolution. This approach fosters collaboration between data scientists and astronomers, driving innovation in methodologies and contributing to significant scientific discoveries. \ShortName~not only serves as a practical tool for current research but also lays the groundwork for future advancements,

\section{\textbf{Summary}}

This study presents \ShortName, a cutting-edge deep learning model designed to revolutionize galactic spectral analysis by providing rapid, accurate estimates of stellar age, metallicity, and dust attenuation. Leveraging an expanded E-MILES dataset with 111,936 varied templates, \ShortName~is trained to disentangle complex spectral features, incorporating multiple dust attenuation and noise levels to enhance robustness. Its architecture, built on four parallel attention-based encoders with multi-scale kernels, efficiently predicts a high-resolution $12 \times 53$ metallicity-age grid — significantly finer than the grids typically used in spectral fitting tools — setting a new benchmark in spectral analysis. Performance evaluation on both synthetic and real galactic spectra demonstrates very good accuracy, with spectral fit residuals averaging below 0.3\% and a standard deviation of around 5\%. However, its true breakthrough lies in computational efficiency: \ShortName~accelerates spectral fitting from 11 seconds per spectrum with pPXF to under 4 milliseconds—a staggering speed-up of over 2,750 times. Beyond speed, it is also 68 times more power-efficient, significantly reducing the computational footprint for large-scale astronomical surveys. While currently optimized for PHANGS-MUSE data, \ShortName's adaptable framework allows fine-tuning for different instruments, spectral resolutions, and observational conditions, ensuring its versatility across future surveys. By combining state-of-the-art deep learning, computational sustainability, and large-scale scalability, \ShortName~offers a transformative approach to mapping galactic properties, paving the way for faster, more efficient, and environmentally conscious astronomical research.

\section{\textbf{Acknowledgements}}
We would like to extend our sincere thanks to the PHANGS-MUSE team for generously sharing the spectral data and maps for NGC4254 and NGC5068. Their contribution enabled the detailed analysis and validation of \ShortName's performance. 
Luca Cortese and Adam B. Watts acknowledge support from the Australian Research Council Discovery Project funding scheme (DP210100337).

\bibliographystyle{apalike} 
\bibliography{references}  

\begin{thebibliography}{}

\bibitem[{Bacon} et~al., 2017]{2017A&A...608A...1B}
{Bacon}, R., {Conseil}, S., {Mary}, D., {Brinchmann}, J., {Shepherd}, M., {Akhlaghi}, M., {Weilbacher}, P.~M., {Piqueras}, L., {Wisotzki}, L., {Lagattuta}, D., {Epinat}, B., {Guerou}, A., {Inami}, H., {Cantalupo}, S., {Courbot}, J.~B., {Contini}, T., {Richard}, J., {Maseda}, M., {Bouwens}, R., {Bouch{\'e}}, N., {Kollatschny}, W., {Schaye}, J., {Marino}, R.~A., {Pello}, R., {Herenz}, C., {Guiderdoni}, B., and {Carollo}, M. (2017).
\newblock {The MUSE Hubble Ultra Deep Field Survey. I. Survey description, data reduction, and source detection}.
\newblock {\em \aap}, 608:A1.

\bibitem[Cappellari and Emsellem, 2004]{cappellari2004parametric}
Cappellari, M. and Emsellem, E. (2004).
\newblock Parametric recovery of line-of-sight velocity distributions from absorption-line spectra of galaxies via penalized likelihood.
\newblock {\em Publications of the Astronomical Society of the Pacific}, 116(816):138.

\bibitem[Carnall et~al., 2018]{carnall2018}
Carnall, A., McLure, R., Dunlop, J., and Dav{\'e}, R. (2018).
\newblock Inferring the star formation histories of massive quiescent galaxies with bagpipes: evidence for multiple quenching mechanisms.
\newblock {\em Monthly Notices of the Royal Astronomical Society}, 480(4):4379--4401.

\bibitem[Chabrier, 2001]{chabrier2001galactic}
Chabrier, G. (2001).
\newblock The galactic disk mass budget. i. stellar mass function and density.
\newblock {\em The Astrophysical Journal}, 554(2):1274.

\bibitem[{Emsellem} et~al., 2022]{2022A&A...659A.191E}
{Emsellem}, E., {Schinnerer}, E., {Santoro}, F., {Belfiore}, F., {Pessa}, I., {McElroy}, R., {Blanc}, G.~A., {Congiu}, E., {Groves}, B., {Ho}, I.~T., {Kreckel}, K., {Razza}, A., {Sanchez-Blazquez}, P., {Egorov}, O., {Faesi}, C., {Klessen}, R.~S., {Leroy}, A.~K., {Meidt}, S., {Querejeta}, M., {Rosolowsky}, E., {Scheuermann}, F., {Anand}, G.~S., {Barnes}, A.~T., {Be{\v{s}}li{\'c}}, I., {Bigiel}, F., {Boquien}, M., {Cao}, Y., {Chevance}, M., {Dale}, D.~A., {Eibensteiner}, C., {Glover}, S. C.~O., {Grasha}, K., {Henshaw}, J.~D., {Hughes}, A., {Koch}, E.~W., {Kruijssen}, J.~M.~D., {Lee}, J., {Liu}, D., {Pan}, H.-A., {Pety}, J., {Saito}, T., {Sandstrom}, K.~M., {Schruba}, A., {Sun}, J., {Thilker}, D.~A., {Usero}, A., {Watkins}, E.~J., and {Williams}, T.~G. (2022).
\newblock {The PHANGS-MUSE survey. Probing the chemo-dynamical evolution of disc galaxies}.
\newblock {\em \aap}, 659:A191.

\bibitem[Fabbro et~al., 2018]{fabbro2018starnet}
Fabbro, S., Venn, K., O'Briain, T., Bialek, S., Kielty, C., Jahandar, F., and Monty, S. (2018).
\newblock An application of deep learning in the analysis of stellar spectra.
\newblock {\em Monthly Notices of the Royal Astronomical Society}, 475(3):2978--2993.

\bibitem[Falc{\'o}n-Barroso et~al., 2011]{falcon2011miles}
Falc{\'o}n-Barroso, J., S{\'a}nchez-Bl{\'a}zquez, P., Vazdekis, A., Ricciardelli, E., Cardiel, N., Cenarro, A., Gorgas, J., and Peletier, R. (2011).
\newblock An updated miles stellar library and stellar population models.
\newblock {\em Astronomy \& Astrophysics}, 532:A95.

\bibitem[Lovell et~al., 2019]{lovell2019}
Lovell, C.~C., Acquaviva, V., Thomas, P.~A., Iyer, K.~G., Gawiser, E., and Wilkins, S.~M. (2019).
\newblock Learning the relationship between galaxies spectra and their star formation histories using convolutional neural networks and cosmological simulations.
\newblock {\em Monthly Notices of the Royal Astronomical Society}, 490(4):5503--5520.

\bibitem[Neumann et~al., 2021]{neumann2021sdss}
Neumann, J., Thomas, D., Maraston, C., Goddard, D., Lian, J., Hill, L., Dom{\'\i}nguez~S{\'a}nchez, H., Bernardi, M., Margalef-Bentabol, B., Barrera-Ballesteros, J.~K., et~al. (2021).
\newblock Sdss-iv manga: drivers of stellar metallicity in nearby galaxies.
\newblock {\em Monthly Notices of the Royal Astronomical Society}, 508(4):4844--4857.

\bibitem[O'Donnell, 1994]{o1994rnu}
O'Donnell, J.~E. (1994).
\newblock Rnu-dependent optical and near-ultraviolet extinction.
\newblock {\em Astrophysical Journal, Part 1 (ISSN 0004-637X), vol. 422, no. 1, p. 158-163}, 422:158--163.

\bibitem[Pasquet et~al., 2019]{pasquet2019}
Pasquet, J., Bertin, E., Treyer, M., Arnouts, S., and Fouchez, D. (2019).
\newblock Photometric redshifts from sdss images using a convolutional neural network.
\newblock {\em Astronomy \& Astrophysics}, 621:A26.

\bibitem[Pessa et~al., 2023]{pessa2023resolved}
Pessa, I., Schinnerer, E., Sanchez-Blazquez, P., Belfiore, F., Groves, B., Emsellem, E., Neumann, J., Leroy, A., Bigiel, F., Chevance, M., et~al. (2023).
\newblock Resolved stellar population properties of phangs-muse galaxies.
\newblock {\em Astronomy \& Astrophysics}, 673:A147.

\bibitem[S{\'a}nchez-Bl{\'a}zquez et~al., 2006]{sanchez2006medium}
S{\'a}nchez-Bl{\'a}zquez, P., Peletier, R., Jim{\'e}nez-Vicente, J., Cardiel, N., Cenarro, A.~J., Falcon-Barroso, J., Gorgas, J., Selam, S., and Vazdekis, A. (2006).
\newblock Medium-resolution isaac newton telescope library of empirical spectra.
\newblock {\em Monthly Notices of the Royal Astronomical Society}, 371(2):703--718.

\bibitem[van~de Sande et~al., 2022]{van2022geckos}
van~de Sande, J., Fraser-McKelvie, A., Fisher, D., Martig, M., Hayden, M., et~al. (2022).
\newblock Geckos: Turning galaxy evolution on its side with deep observations of edge-on galaxies.
\newblock {\em Proceedings of the International Astronomical Union}, 18(S377):27--33.

\bibitem[Vaswani et~al., 2017]{vaswani2017attention}
Vaswani, A., Shazeer, N., Parmar, N., Uszkoreit, J., Jones, L., Gomez, A.~N., Kaiser, {\L}., and Polosukhin, I. (2017).
\newblock Attention is all you need.
\newblock {\em Advances in neural information processing systems}, 30.

\bibitem[Vazdekis et~al., 2016]{vazdekis2016uv}
Vazdekis, A., Koleva, M., Ricciardelli, E., R{\"o}ck, B., and Falc{\'o}n-Barroso, J. (2016).
\newblock Uv-extended e-miles stellar population models: young components in massive early-type galaxies.
\newblock {\em Monthly Notices of the Royal Astronomical Society}, 463(4):3409--3436.

\bibitem[Vazdekis et~al., 2010]{vazdekis2010miles}
Vazdekis, A., S{\'a}nchez-Bl{\'a}zquez, P., Falc{\'o}n-Barroso, J., Cenarro, A., Beasley, M., Cardiel, N., Gorgas, J., and Peletier, R. (2010).
\newblock Evolutionary stellar population synthesis with miles--i. the base models and a new line index system.
\newblock {\em Monthly Notices of the Royal Astronomical Society}, 404(4):1639--1671.

\bibitem[{Watts} et~al., 2024]{2024MNRAS.530.1968W}
{Watts}, A.~B., {Cortese}, L., {Catinella}, B., {Fraser-McKelvie}, A., {Emsellem}, E., {Coccato}, L., {van de Sande}, J., {Brown}, T.~H., {Ascasibar}, Y., {Battisti}, A., {Boselli}, A., {Davis}, T.~A., {Groves}, B., and {Thater}, S. (2024).
\newblock {MAUVE: a 6 kpc bipolar outflow launched from NGC 4383, one of the most H I-rich galaxies in the Virgo cluster}.
\newblock {\em \mnras}, 530(2):1968--1983.

\bibitem[Woo et~al., 2024]{woo2024comparison}
Woo, J., Walters, D., Archinuk, F., Faber, S., Ellison, S.~L., Teimoorinia, H., and Iyer, K. (2024).
\newblock Stellar populations with optical spectra: deep learning versus popular spectrum fitting codes.
\newblock {\em Monthly Notices of the Royal Astronomical Society}, 530(4):4260--4276.

\end{thebibliography}

\end{document}